\documentclass[aps,prb,twocolumn,longbibliography] {revtex4-2}

\usepackage{amsmath,graphicx,amsfonts}

\usepackage{times}
\usepackage[colorlinks=true,linkcolor=blue,urlcolor=blue,citecolor=blue, breaklinks=true,hypertexnames=false]{hyperref}

\usepackage{amssymb}
\usepackage{bbm}

\usepackage[utf8]{inputenc}
\DeclareUnicodeCharacter{03BC}{\mbox{$\mu$}}
\DeclareUnicodeCharacter{03B4}{$\delta$}
\DeclareUnicodeCharacter{2009}{-}

\usepackage[english]{babel}
\addto\captionsenglish{}
\addto\captionsenglish{}

\makeatletter
\def\bbl@set@language#1{%
	\edef\languagename{%
		\ifnum\escapechar=\expandafter`\string#1\@empty
		\else\string#1\@empty\fi}%
	\@ifundefined{babel@language@alias@\languagename}{}{%
		\edef\languagename{\@nameuse{babel@language@alias@\languagename}}%
	}%
	\select@language{\languagename}%
	\expandafter\ifx\csname date\languagename\endcsname\relax\else
	\if@filesw
	\protected@write\@auxout{}{\string\select@language{\languagename}}%
	\bbl@for\bbl@tempa\BabelContentsFiles{%
		\addtocontents{\bbl@tempa}{\xstring\select@language{\languagename}}}%
	\bbl@usehooks{write}{}%
	\fi
	\fi}
\newcommand{\DeclareLanguageAlias}[2]{%
	\global\@namedef{babel@language@alias@#1}{#2}%
}
\makeatother

\DeclareLanguageAlias{en}{english}

\makeatletter
\def\@bibdataout@aps{%
	\immediate\write\@bibdataout{%
		@CONTROL{%
			apsrev41Control%
			\longbibliography@sw{%
				,author="08",editor="1",pages="0",title="1",year="1"%
			}{%pages="0" means no abcd in 123-abcd, so only the initial page appears. pages="1" means 123-abcd will appear. title="" means no titles.
			}%
		}%
	}%
\if@filesw \immediate \write \@auxout {\string \citation {apsrev41Control}}\fi 
}
\makeatother

\def\F {\scriptscriptstyle{F}}
\def\B {\scriptscriptstyle{B}}

\def\Q {\scriptscriptstyle{Q}}

\def\BB {\scriptscriptstyle{B\!B}}
\def\FF {\scriptscriptstyle{F\!F}}
\def\BF {\scriptscriptstyle{B\!F}}
\def\FB {\scriptscriptstyle{F\!B}}

\date\today

\begin{document}
\title{Bethe-ansatz study of the Bose-Fermi mixture}
\author{Soham Chandak}
\author{Aleksandra Petkovi\'{c}}
\author{Zoran Ristivojevic}
\affiliation{Universit\'{e} de Toulouse, CNRS, Laboratoire de Physique Th\'{e}orique, Toulouse, France}
	
\begin{abstract}
We consider a one-dimensional mixture of bosons and spinless fermions with contact interactions. In this system, the elementary excitations at low energies are described by four linearly dispersing modes characterized by two excitation velocities. Here we study the velocities in a system with equal interaction strengths and equal masses of bosons and fermions. The resulting model is integrable and admits an exact Bethe-ansatz solution. We analyze it and analytically derive various exact results, which include the Drude weight matrix. We show that the excitation velocities can be calculated from the knowledge of the matrices of compressibility and the Drude weights, as their squares are the eigenvalues of the product of the two matrices. The elements of the Drude weight matrix obey certain sum rules as a consequence of Galilean invariance. Our results are consistent with the presence of a momentum-momentum coupling term between the two subsystems of bosons and fermions in the effective low-energy Hamiltonian. The analytical method developed in the present study can be extended to other models that possess a nested Bethe-ansatz structure.
\end{abstract}
\maketitle
	
\section{Introduction}\label{sec1}
	
The low-energy properties of one-dimensional quantum liquids are described by the Luttinger liquid  \cite{haldane_effective_1981}. The excitations in this theory have a linear spectrum and represent the waves of the particle density. The Luttinger liquid is characterized by two parameters, $v$ and $K$. The former denotes the velocity of waves, and the latter controls the decay of various correlation functions at long distances. From the microscopic point of view, one of the primary theoretical goals is the calculation of $v$ and $K$ knowing the interaction potential between the particles.
	
The latter problem is generally complicated in the case of interactions of arbitrary strengths \cite{giamarchi}. It can, however, be simplified by resorting to various relations between $v$, $K$, and other physical quantities. One such relation arises in Galilean-invariant systems, where $v$ and $K$ are not independent. They obey the constraint $m v K= \pi \hbar n$, where $m$ denotes the mass of particles and $n$ is their mean density \cite{haldane_effective_1981}. Another connection originates from the phenomenological thermodynamics. The relation between the compressibility and the sound velocity in this approach can be expressed as
\begin{align}\label{eq:v2mu(n)}
v^2=\frac{n}{m}\frac{\partial \mu}{\partial n},
\end{align}
where $\mu$ is the chemical potential. Thus the density dependence of the chemical potential $\mu(n)$ fully determines the sound velocity and the Luttinger liquid parameter.
	
In the emerging phenomenological Luttinger liquid character of one-dimensional liquids we plausibly assumed that the sound velocity coincides with the velocity of excitations. This scenario was initially proved by Lieb \cite{lieb_exact_1963b} in a microscopic theory of the nonrelativistic Bose gas with contact repulsions \cite{lieb_exact_1963a}. In a more general study, Haldane \cite{haldane_demonstration_1981} showed that the Luttinger liquid character holds for a wider class of models solvable by the Bethe ansatz, the Lieb--Liniger model of the Bose gas being a particular example. Therefore, Bethe-ansatz solvable models are the prime examples where the parameters of the Luttinger liquid can be calculated analytically. Nevertheless, this is still a challenging task, as the analysis of the Bethe ansatz is often complicated beyond a numerical treatment.
	
The previous considerations apply to one-component quantum liquids. Analogous question about the low-energy excitations can be posed for multicomponent liquids. The simplest example is given by repulsive spin-$\frac{1}{2}$ fermions. The diagonalization of its Hamiltonian reveals that the excitations are characterized by two velocities corresponding to charge and spin degrees of freedom. The latter is known as a spin-charge separation as the velocities are distinct \cite{giamarchi}. Similar picture also arises in other multicomponent  liquids such as the Bose-Fermi mixture. This system was studied using the phenomenological Luttinger liquid approach in various works, see, for example, Refs.~\cite{cazalilla_2003,wang_2007,orignac_2010}.

The problem of multicomponent quantum liquids, on the other hand, can be studied using a microscopic approach based on the Bethe ansatz, which leads to formally exact solutions. Unlike the Lieb--Liniger model, for example, where the structure of the obtained equations for the wave function is relatively simple, in the multicomponent case one has to deal with the so-called nested Bethe ansatz \cite{yang_exact_1967,gaudin}. The resulting equations are coupled and therefore more complicated for analysis. The examples are the Hubbard model \cite{lieb_absence_1968} and various mixtures consisting of multicomponent bosons and fermions with equal masses \cite{sutherland,sutherland_gen}. In particular cases, the latter include spin-$\frac{1}{2}$ fermions \cite{yang_exact_1967,gaudin_exact_1967}, a mixture of bosons and spin-$\frac{1}{2}$ fermions \cite{lai_yang_exact}, as well as a mixture of bosons and spinless fermions \cite{lai_yang_exact,imambekov_exactly_2006,batchelor_2005}. 

In this paper we study the latter Bose-Fermi mixture. Although the exact solution has been known for a long time \cite{lai_yang_exact,lai_thermodynamics_1974}, the properties of this system remained largely unexplored till the interest was revived in Ref.~\cite{imambekov_exactly_2006}. Since then various properties of the model have been studied using the Bethe ansatz such as the ground-state energy \cite{imambekov_exactly_2006,imambekov_applications_2006, batchelor_2005}, the correlation functions \cite{imambekov_applications_2006,palacios_2005,patu_universal_2017}, and the thermodynamics \cite{guan_2012,patu_momentum_2019}. The excitation velocities were also studied in the limiting cases of weak and strong interactions \cite{batchelor_2005}. One of our goals here is to understand the excitation velocities from a more general point of view and connect them with some thermodynamic quantities. In the case of one-component Galilean-invariant liquids, this is achieved by Eq.~(\ref{eq:v2mu(n)}), which connects the velocity and the compressibility. In the present case it is not obvious how to connect the compressibility matrix and the two velocities. Unlike the one-component liquid case, we find that the knowledge of the density dependence of the chemical potentials of the two constituent particle species is not sufficient in order to determine the excitation velocities. The necessary additional information can be obtained from the Drude weight matrix. It measures the response of the system to the twisted boundary conditions. In the one-component case, the Drude weight is proportional to the particle density \cite{sutherland_adiabatic_1990}. It can thus be understood as present in Eq.~(\ref{eq:v2mu(n)}), although it was not explicitly stated. We find that the squares of the excitation velocities are the eigenvalues of the matrix that equals  the product of the compressibility and the Drude weight matrices. This result is consistent with the simple case expressed by Eq.~(\ref{eq:v2mu(n)}) \footnote{We loosely call $\partial \mu/\partial n$ the compressibility, similar as in Ref.~\cite{imambekov_exactly_2006}, while in reality, $\partial \mu/\partial n$ is proportional to the inverse compressibility. We keep the same logic for the compressibility matrix.}. In addition, we derive microscopically exact results for the matrix elements of the compressibility and the Drude weight matrices. We note that the Drude weights were earlier studied in other integrable models, see for example Refs.~\cite{shastry_twisted_1990,sutherland_adiabatic_1990,fujimoto_exact_1998,zotos_finite_1999,benz_finite_2005,sirker_diffusion_2009,doyon_drude_2017,luo_quantum_2025,gohmann_ballistic_2025}, but not in the Bose-Fermi mixture.

The outline of this work is as follows. In Sec.~\ref{sec:exactsolution} we introduce the model of the Bose-Fermi mixture with contact repulsions. We discuss its Bethe-ansatz solution, in particular the ground state, the excitations, as well as the excitation velocities. In Sec.~\ref{sec:compressibility} we introduce the compressibility matrix and derive its decomposition that involves the excitation velocities and the matrix elements of the dressed charge matrix. In Sec.~\ref{sec:additionalrelations} we derive the relations due to Galilean invariance of the system that involve the velocities, the densities, and the matrix elements of the dressed charge matrix. In Sec.~\ref{sec:Drudeweight} we study the system under the twisted boundary conditions and derive the Drude weight matrix. In Sec.~\ref{sec:velocities} we obtain the general relation that connects  the velocities of excitations, the Drude weight, and the compressibility matrices. The summary of our work is given in Sec.~\ref{sec:discussions}. Various technical details are explained in Appendixes \ref{sec:reductiontoLL}--\ref{appendix:last}.

\section{Formal exact solution}\label{sec:exactsolution}

\subsection{Hamiltonian}

We study an interacting one-dimensional mixture of $N_{\B}$ bosons and $N_{\F}$ spinless fermions. We assume that the bosons interact among themselves and with fermions, while the fermions do not interact among themselves. In the case of contact interactions between all the species, the resulting Hamiltonian is given by
\begin{align}\label{eq:Hamiltonian}
	H = {}& \frac{\hbar^{2}}{2m}\Biggl[- \sum_{j=1}^{N}\frac{\partial^2}{\partial x_{j}^2} +c \sum_{j,l=1\atop j\neq l}^{N}\delta(x_{j}-x_{l})\Biggr].
\end{align}
Here $c > 0$ denotes the repulsion strength, $m$ is the mass of particles, and  $N = N_{\B}+N_{\F}$ is the total number of particles. The wave function of the system $\Psi(x_1,\ldots,x_{N_{{\B}}},x_{N_{{\B}}+1},\ldots,x_N)$ depends on the coordinates of bosons and fermions, and satisfies the eigenvalue problem $H\Psi=E\Psi$, where $E$ is the eigenenergy. The wave function is symmetric with respect to the permutations of bosonic coordinates and antisymmetric with respect to the permutations of fermionic ones. The wave function thus vanishes if two fermionic coordinates are identical, which is the manifestation of the Pauli principle. Therefore, the $\delta$-function interaction in the Hamiltonian (\ref{eq:Hamiltonian}) that formally exists between any two particles, does not affect the wave function if the particles are the fermions. Otherwise, the $\delta$ function describes the repulsion between two particles when they share the same position.

\subsection{Bethe-ansatz solution}

In the Hamiltonian (\ref{eq:Hamiltonian}) we assumed equal masses of bosons and fermions and equal interaction strengths between boson-boson and boson-fermion pairs. In this case the model is exactly solvable in terms of the Bethe ansatz \cite{lai_yang_exact,imambekov_applications_2006}. The corresponding wave function of the system is parametrized by the set of quasimomenta (or rapidities) ${k_j}$, where $j=1,2,\ldots,N$. In the case of a system with periodic boundary conditions, the quasimomenta are determined by the set of discrete Bethe-ansatz equations 
\begin{subequations}
\label{eqn:BA_discrete}
\begin{align}
\label{eqn:BA_discrete_1}
e^{iL k_{j}} ={}& \prod_{l=1}^{N_{\B}}\frac{k_{j}-\Lambda_{l}+\frac{i c}{2}}{k_{j}-\Lambda_{l}-\frac{i c}{2}}, \quad j = 1,2,\ldots,N, \\
\label{eqn:BA_discrete_2}
1 = {}&\prod_{j=1}^{N}\frac{k_{j}-\Lambda_{l}+\frac{i c}{2}}{k_{j}-\Lambda_{l}-\frac{i c}{2}}, \quad l = 1,2,\ldots,N_{\B}.
\end{align}
\end{subequations}
Here $L$ is the system size. Equations (\ref{eqn:BA_discrete}) contain a set of auxiliary quasimomenta $\Lambda_l$ that participate in the expressions that determine the quasimomenta $k_{j}$ but do not explicitly enter the wave function. 
Note that the quasimomenta $k_j$ and $\Lambda_l$ that obey Eqs.~(\ref{eqn:BA_discrete}) are real~\cite{imambekov_applications_2006}.

Instead of using the initial form (\ref{eqn:BA_discrete}), after taking the logarithm we can express the Bethe-ansatz equations as
\begin{subequations}\label{eqn:DBAElog}
\begin{gather}
Lk_{j} = 2\pi I_{j} + \sum_{l=1}^{N_{\B}}\theta(2k_{j}-2\Lambda_{l}),\quad j = 1,2,\ldots,N,\\
2\pi J_{l} = \sum_{j=1}^{N}\theta(2k_{j}-2\Lambda_{l}),\quad l = 1,2,\ldots,N_{\B}.
\end{gather}
\end{subequations}
Here we have introduced the two-particle scattering phase shift
\begin{align}\label{eq:phaseshift}
	\theta(k)=-2\arctan(k/c),  
\end{align}
while $I_{j}$ and $J_{l}$ are the quantum numbers describing the state of the system. They take integer or odd half-integer values depending on the parity of $N$ and $N_{\B}$. The momentum and the energy of the system corresponding to the Bethe wave function are 
\begin{align}\label{eq:pe}
p = \hbar\sum_{j=1}^{N} k_{j},\quad E = \frac{\hbar^2}{2m} \sum_{j=1}^{N}k_{j}^{2}.
\end{align}
Equations (\ref{eqn:DBAElog}) describe any eigenfunction of the Hamiltonian that has the symmetry specified below Eq.~(\ref{eq:Hamiltonian}). 

The ground state of the system is realized for the quantum numbers that are consecutive integers or odd half-integers symmetrically distributed around zero. They are given by \cite{imambekov_applications_2006}
\begin{subequations}\label{eqn:quantumnumbers}
\begin{gather}
I_{j} = j - \frac{N+1}{2}, \quad j = 1,2,\ldots,N,\\
J_{l} = l - \frac{N_{\B}+1}{2}, \quad l = 1,2,\ldots,N_{\B}.
\end{gather}    
\end{subequations}
Due to the periodic boundary conditions imposed when deriving Eqs.~(\ref{eqn:BA_discrete}), we are restricted to study the system with odd number of fermions~\footnote{One way to see it is to consider the case $N_{\B}=0$. Then $k_j=2\pi I_j/L$ are compatible with momenta of free fermions subject to the periodic (antiperiodic) boundary conditions for odd (even) $N_{\F}$.}. In this case $N$ and $N_{\B}$ are of different parity. Indeed, once Eqs.~(\ref{eqn:DBAElog}) are solved for the choice of quantum numbers (\ref{eqn:quantumnumbers}), we verified that Eqs.~(\ref{eqn:BA_discrete}) are satisfied if $N$ and $N_{\B}$ are of different parity. This is nevertheless not a very important restriction for our study in the thermodynamic limit.

Excited states of the system are described by the wave functions characterized by choices of the quantum numbers different from the one of Eq.~(\ref{eqn:quantumnumbers}). Among many excitations, we can distinguish four types of elementary ones, in analogy to Lieb's classification for the Lieb--Liniger model \cite{lieb_exact_1963b}. Any excitation of the system can be expressed as a combination of these elementary excitations. There are type-I and type-II excitations corresponding to particular perturbations of the ground-state quantum numbers for $I_j$ and $J_l$. In particular, a type-I excitation for $I_j$ corresponds to the promotion of the highest quantum number $I_N$ to a value larger than it \footnote{This corresponds to a right-moving excitation. There is analogous left-moving one that corresponds to a promotion of $I_1$ to a smaller value. Since this symmetry is not very important in the following, we only consider the excitations that are obtained by increasing the ground-state quantum numbers.}. Similarly, a type-I excitation for $J_l$ corresponds to a promotion of the highest quantum number $J_{N_{\B}}$ to a larger value. A type-II excitation for $I_j$ corresponds to a promotion of any ground-state $I$ quantum number to the value $(N+1)/2$, which is the quantum number of the first unoccupied state above the ground-state sea. Similarly, a type-II excitation for $J_l$ corresponds to a promotion of any ground-state $J$ quantum number to the value $(N_{\B}+1)/2$.
The energy and momentum of elementary excitations are determined by Eq.~(\ref{eq:pe}).

\subsection{Thermodynamic limit}

Instead of treating the discrete Bethe-ansatz equations, we consider the thermodynamic limit where the system size $L$, the total number of particles $N$, and the number of bosons $N_{\B}$ tend to infinity such that the particle density $n=N/L$ and the density of bosons $n_{\B}=N_{\B}/L$ are kept fixed. In this case the density of fermions $n_{\F}=n-n_{\B}$ is also fixed. In the thermodynamic limit, the quasimomenta $k_j$ and $\Lambda_l$ become continuous. We characterize them by the density functions  $\rho(k)$ and $\sigma(\Lambda)$ introduced in such a way that 
\begin{subequations}\label{eq:denden}
\begin{gather}
L\rho(k)d k = \textrm{number of $k$'s in } [k,k+dk],\\
L\sigma(\Lambda)d\Lambda = \textrm{number of $\Lambda$'s in } [\Lambda,\Lambda+d \Lambda].
\end{gather}
\end{subequations}
In the ground state, the quantum numbers $I_j$ and $J_l$ are consecutive. The corresponding quasimomenta will be compactly spread between the two so-called Fermi rapidities; these are $-Q$ and $Q$ for $k$'s and $-B$ and $B$ for $\Lambda$'s. Subtracting the discrete equations (\ref{eqn:DBAElog}) with indices $j$ from the same equations with $j+1$, after using the definitions (\ref{eq:denden}), in the continuum we obtain the set of coupled integral equations
\begin{subequations}
\label{eqn:Ground_state_root}
\begin{gather}
\label{eqn:Ground_state_root_1}
\rho(k) + \frac{1}{\pi}\int_{-B}^{B}dq \;\! \theta'(2k-2q) \sigma(q) = \frac{1}{2\pi},\\
\label{eqn:Ground_state_root_2}
\sigma(k) + \frac{1}{\pi}\int_{-Q}^{Q}dq \;\! \theta'(2k-2q) \rho(q) = 0.
\end{gather}
\end{subequations}
They determine the densities $\rho(k)$ and $\sigma(k)$. The kernel in Eqs.~(\ref{eqn:Ground_state_root}) is given by the derivative of the phase shift,
\begin{align}\label{eq:kernel}
 \theta'(k)=-\frac{2c}{c^2+k^2}.
\end{align}
The densities of quasimomenta determine the particle densities via
\begin{align}
\label{eqn:GS_density}
	n=\int_{-Q}^{Q}dk\;\!\rho(k), \quad n_{\B}=\int_{-B}^{B}dk\;\!\sigma(k),
\end{align}
as well as the ground-state energy,
\begin{align}\label{eq:E0}
	E_0 = \frac{\hbar^2 L}{2m}\int_{-Q}^{Q}dk\;\! k^2\rho(k).
\end{align}
Note that the functions $\rho(k)$ and $\sigma(k)$ defined by Eqs.~(\ref{eqn:Ground_state_root}) depend in fact on two more variables, $Q$ and $B$. Therefore it would have been more accurate to use $\rho(k,Q,B)$ and $\sigma(k,Q,B)$ from the outset. Then, e.g., in Eqs.~(\ref{eqn:GS_density}) and (\ref{eq:E0}) we would need $n(Q,B)$, $n_{\B}(Q,B)$, and $E_0(Q,B)$. However, for easier subsequent notation the simplified form $\rho(k)$, $\sigma(k)$ is kept in many expressions and, where needed, longer notation will be used. We note that Eqs.~(\ref{eqn:Ground_state_root}) in the case $B\to\infty$ reduce to the Lieb--Liniger model, as we discuss in Appendix~\ref{sec:reductiontoLL}.

The momentum and the energy of elementary excitations can also be studied in the thermodynamic limit using the standard methods \cite{korepin,takahashi}. They are encoded in the functions $\xi(k)$ and $\omega(k)$ that are defined in terms of the coupled integral equations
\begin{subequations}
\label{eqn:BF_exc_spec}
\begin{gather}\label{eqn:BF_exc_spec_1}
\xi(k)+ \frac{1}{\pi}\int_{-B}^{B}dq \;\! \theta'(2k-2q) \omega(q) = \frac{\hbar^2 k}{m},\\
\label{eqn:BF_exc_spec_2}
\omega(k) + \frac{1}{\pi}\int_{-Q}^{Q}dq \;\! \theta'(2k-2q) \xi(q) = 0.
\end{gather}
\end{subequations}
The momentum and the energy of right-moving type-I and type-II excitations obtained by perturbing $I_j$ quantum numbers are given by 
\begin{align}\label{eq:typeI-Ij}
	p = 2\pi\hbar\left |\int_{Q}^{\tau}dk\rho(k)\right|, \quad \varepsilon =  \left|\int_{Q}^{\tau}dk\;\! \xi(k)\right|.
\end{align}
Here $\tau$ is a parameter that defines $p$ and $\varepsilon$. For type-I excitations, $\tau>Q$; for type-II excitations, $-Q\le \tau<Q$. The momentum and the energy of type-I and type-II excitations in the quantum numbers $J_{l}$ are given by 
\begin{align}\label{eq:typeI-Jl}
	p = 2\pi\hbar\left|\int_{B}^{\tau}dk\;\! \sigma(k)\right|, \quad \varepsilon = \left|  \int_{B}^{\tau} dk\;\!\omega(k)\right|.
\end{align}
Here $\tau>B$ for type-I excitations, while $-B\le\tau<B$ for type-II ones. 

We have found four different types of elementary excitations. They are linear functions of the momenta when the momenta are small. Moreover, type-I and type-II branches for the perturbations of the same quantum numbers are characterized by the same velocity that is defined by 
\begin{align}\label{eq:vvvv}
	v = \left.\frac{\partial \varepsilon}{\partial p}\right|_{p\to 0}.
\end{align}
There are thus two distinct excitation velocities in our system corresponding to the excitations in the two sets of quantum numbers ${I_{j}}$ and ${J_{l}}$. Evaluating Eq.~(\ref{eq:vvvv}) by parametric differentiation of Eqs.~(\ref{eq:typeI-Ij}) and (\ref{eq:typeI-Jl}), the velocities are, respectively, given by 
\begin{align}\label{eq:velocities}
	v_{1} = \frac{1}{2\pi\hbar}\frac{\xi(Q)}{\rho(Q)}, \quad 
	v_{2} = \frac{1}{2\pi\hbar}\frac{\omega(B)}{\sigma(B)}.  
\end{align}
Note that $v_1$ and $v_2$ defined by Eq.~(\ref{eq:velocities}) are nonzero for finite $c$, $B$, and $Q$. Moreover, the velocities are positive, which is expected for right-moving excitations; a mathematical proof is given in Appendix \ref{appendix-functional}. One of our goals in the following is to express the velocities (\ref{eq:velocities}) in terms of the thermodynamic quantities.

\section{The compressibility matrix}\label{sec:compressibility}

In this section we study the compressibility matrix
\begin{align}
	\mathcal{M}=\begin{pmatrix}
		\mathcal{M}_{\FF} & \mathcal{M}_{\FB}\\
		\mathcal{M}_{\BF} & \mathcal{M}_{\BB}
	\end{pmatrix},
\end{align}
where the matrix elements are conveniently defined by
\begin{align}
\mathcal{M}_{ss'} = \frac{1}{\pi\hbar}\frac{\partial \mu_{s}}{\partial n_{s'}},\quad s,s'\in \{F,B\}.
\end{align}
Here $\mu_{s}$ denotes the chemical potential of the species $s$. The latter are given by the standard expressions
\begin{align}\label{eq:chemicalpotentials}
\mu_{\F} = \frac{\partial E_{0}}{\partial N_{\F}}, \quad \mu_{\B} = \frac{\partial E_{0}}{\partial N_{\B}},
\end{align}
where $E_0$ is the ground-state energy. The chemical potentials are the functions of $n_{\F}$ and $n_{\B}$, and can be understood as composite functions $\mu_s\boldsymbol{(}Q(n_{\F},n_{\B}),B(n_{\F},n_{\B})\boldsymbol{)}$. Therefore, using the chain rule, we can split the compressibility matrix as 
\begin{align}
	\label{eqn:Matrix_split}
	\mathcal{M} = \frac{1}{\pi\hbar}\begin{pmatrix}
		\frac{\partial \mu_{\!_F}}{\partial Q} & \ \frac{\partial \mu_{\!_F}}{\partial B}  \\
		\frac{\partial \mu_{\!_B}}{\partial Q} &\ \frac{\partial \mu_{\!_B}}{\partial B} 
	\end{pmatrix} \begin{pmatrix}
		\frac{\partial Q}{\partial n_{\!_F}} & \frac{\partial Q}{\partial n_{\!_B}}\\
		\frac{\partial B}{\partial n_{\!_F}} & \frac{\partial B}{\partial n_{\!_B}}
	\end{pmatrix}.
\end{align}
Each of the two latter matrices should be further transformed.

The first matrix on the right-hand side of Eq.~(\ref{eqn:Matrix_split}) can be evaluated from the so-called dressed energy equations \cite{guan_2012} 
\begin{subequations}
\label{eqn:BF_dressed_T0}
\begin{gather}\label{eqn:BF_dressed_T0_1}
\mathcal{E}(k) + \frac{1}{\pi}\int_{-B}^{B}dq\;\!\theta'(2k-2q) \varphi(q) = \frac{\hbar^2k^2}{2m} - \mu_{\F}, \\
\label{eqn:BF_dressed_T0_2}
\varphi(k) + \frac{1}{\pi}\int_{-Q}^{Q} dq\;\!\theta'(2k-2q) \mathcal{E}(q) = \mu_{\F} - \mu_{\B}.
\end{gather}    
\end{subequations}
In these equations, $\mu_{\F}$ and $\mu_{\B}$ are chosen in such a way that for given $Q$ and $B$, the constraints $\mathcal{E}(\pm Q)=0$ and $\varphi(\pm B)=0$ are obeyed. They can thus be seen as the functions $\mu_{\F}(Q,B)$ and $\mu_{\B}(Q,B)$. 
In Appendix \ref{sec:chemical_potential_proof} it is shown that $\mu_{\F}$ and $\mu_{\B}$ are actually the chemical potentials of the system, which are given by Eq.~(\ref{eq:chemicalpotentials}). We note the property
\begin{align}\label{eq:Ephixiomega}
\mathcal{E}'_k(k)=\xi(k),\quad \varphi'_k(k)=\omega(k),
\end{align}
where $\xi(k)$ and $\omega(k)$ are defined by Eqs.~(\ref{eqn:BF_exc_spec}). Here we use the notation $\mathcal{E}'_k(k)=\partial \mathcal{E}(k)/\partial k$.

Differentiating Eqs.~(\ref{eqn:BF_dressed_T0}) with respect to $\mu_{\F}$ we obtain
\begin{gather}\label{eq:dedmuF}
\mathcal{F}\left[\frac{\partial \mathcal{E}}{\partial \mu_{\F}},\frac{\partial \varphi}{\partial \mu_{\F}},B\right] = -1,\quad \mathcal{F}\left[\frac{\partial \varphi}{\partial \mu_{\F}},\frac{\partial \mathcal{E}}{\partial \mu_{\F}},Q\right] = 1.
\end{gather}
Here for convenience we have introduced the functional $\mathcal{F}[\ldots]$ via
\begin{align}\label{eq:Fdefinition}
\mathcal{F}[f,g,B] \equiv f(k) + \frac{1}{\pi}\int_{-B}^{B}dq\;\!\theta'(2k-2q)g(q),
\end{align}
which has some basic properties studied in Appendix \ref{appendix-functional}. Differentiating Eqs.~(\ref{eqn:BF_dressed_T0}) with respect to $\mu_{\B}$ we obtain
\begin{gather}\label{eq:dedmub}
\mathcal{F}\left[\frac{\partial \mathcal{E}}{\partial \mu_{\B}},\frac{\partial \varphi}{\partial \mu_{\B}},B\right] = 0, \quad
\mathcal{F}\left[\frac{\partial \varphi}{\partial \mu_{\B}},\frac{\partial \mathcal{E}}{\partial \mu_{\B}},Q\right] = -1.
\end{gather}
We note that in terms of the functional, the Bethe equations (\ref{eqn:Ground_state_root}) are simply expressed as
\begin{align}\label{eq:BEcompact}
\mathcal{F}[\rho,\sigma,B]=\frac{1}{2\pi},\quad \mathcal{F}[\sigma,\rho,Q]=0.
\end{align}
Equation (\ref{eq:dedmub}) suggests that is is helpful to introduce two new functions $\tilde{\rho}(k)$ and $\tilde{\sigma}(k)$ via the equations
\begin{align}\label{eqn:BF_dual_root}
\mathcal{F}[\tilde{\rho},\tilde{\sigma},Q] = \frac{1}{2\pi}, \quad \mathcal{F}[\tilde{\sigma},\tilde{\rho},B] = 0.
\end{align}
They are related to Eq.~(\ref{eq:BEcompact}) by the interchange of the Fermi rapidities. Using the linearity of the functional $\mathcal{F}$ we find
\begin{subequations}
\label{eqn:dressed_pot_derivatives}
\begin{gather}
		\label{eqn:dressed_pot_derivatives_1}
		\frac{\partial \mathcal{E}(k)}{\partial \mu_{\B}} = -2\pi\tilde{\sigma}(k), \quad \frac{\partial \mathcal{E}(k)}{\partial \mu_{\F}} = -2\pi[\rho(k)-\tilde{\sigma}(k)], \\ 
		\label{eqn:dressed_pot_derivatives_2}
		\frac{\partial \varphi(k)}{\partial \mu_{\B}} = -2\pi\tilde{\rho}(k), \quad \frac{\partial \varphi(k)}{\partial \mu_{\F}} = -2\pi[\sigma(k)-\tilde{\rho}(k)].
\end{gather}
\end{subequations}
Therefore, the derivatives of $\mathcal{E}$ and $\varphi$ that enter Eqs.~(\ref{eq:dedmuF}) and (\ref{eq:dedmub}) are expressed in terms of the densities of rapidities.

The functions $\mathcal{E}$ and $\varphi$ of Eqs.~(\ref{eqn:BF_dressed_T0}) can be understood to have  $\boldsymbol{(}k,\mu_{\F}(Q,B),\mu_{\B}(Q,B)\boldsymbol{)}$ as arguments. Differentiating the nullification constraint $\mathcal{E}(Q)=0$ with respect to $Q$ gives
\begin{align}\label{eqn:mu_der_1}
	\mathcal{E}'_{k}(Q) + \frac{\partial \mathcal{E}(Q)}{\partial \mu_{\B}}\frac{\partial \mu_{\B}}{\partial Q} + \frac{\partial \mathcal{E}(Q)}{\partial \mu_{\F}}\frac{\partial \mu_{\F}}{\partial Q} = 0. 
\end{align}
Similarly, differentiating the same constraint with respect to $B$, we obtain
\begin{align}\label{eqn:mu_der_2}
	\frac{\partial \mathcal{E}(Q)}{\partial \mu_{\B}}\frac{\partial \mu_{\B}}{\partial B} + \frac{\partial \mathcal{E}(Q)}{\partial \mu_{\F}}\frac{\partial \mu_{\F}}{\partial B} = 0.
\end{align}
Analogous operation for the constraint $\varphi(B)=0$ leads to
\begin{align}\label{eqn:mu_der_3}
	\varphi'_{k}(B)+\frac{\partial \varphi(B)}{\partial \mu_{\B}}\frac{\partial \mu_{\B}}{\partial B} + \frac{\partial \varphi(B)}{\partial \mu_{\F}}\frac{\partial \mu_{\F}}{\partial B} = 0
\end{align}
and 
\begin{align}\label{eqn:mu_der_4}
	\frac{\partial \varphi(B)}{\partial \mu_{\B}}\frac{\partial \mu_{\B}}{\partial Q} + \frac{\partial \varphi(B)}{\partial \mu_{\F}}\frac{\partial \mu_{\F}}{\partial Q} = 0.
\end{align}
The latter four expressions in conjunction with the relations (\ref{eqn:dressed_pot_derivatives}) enable us to find
\begin{subequations}
	\begin{align}
		\frac{\partial \mu_{\B}}{\partial Q} &= \frac{\left(\tilde{\rho}_{0}-\sigma_{0}\right)\mathcal{E}'_{k}(Q)}{2\pi(\tilde{\rho}_{0}\rho_{0}-\tilde{\sigma}_{0}\sigma_{0})}, \quad \frac{\partial \mu_{\F}}{\partial Q} = \frac{\tilde{\rho}_{0}\mathcal{E}'_{k}(Q)}{2\pi(\tilde{\rho}_{0}\rho_{0}-\tilde{\sigma}_{0}\sigma_{0})}, \\
		\frac{\partial \mu_{\B}}{\partial B} &= \frac{\left(\rho_{0}-\tilde{\sigma}_{0}\right)\varphi'_{k}(B)}{2\pi(\tilde{\rho}_{0}\rho_{0}-\tilde{\sigma}_{0}\sigma_{0})}, \quad \frac{\partial \mu_{\F}}{\partial B} = -\frac{\tilde{\sigma}_{0}\varphi'_{k}(B)}{2\pi(\tilde{\rho}_{0}\rho_{0}-\tilde{\sigma}_{0}\sigma_{0})}.
	\end{align}
\end{subequations}
Here we have introduced the abbreviations
\begin{subequations}\label{eq:abbreviations}
\begin{gather}
\rho_0=\rho(Q),\quad \sigma_{0}=\sigma(B),\\
\tilde\rho_0=\tilde{\rho}(B),\quad \tilde\sigma_0=\tilde{\sigma}(Q).
\end{gather}
\end{subequations}
Introducing the so-called dressed charge matrix 
\begin{align}\label{eq:Z}
	\mathcal{Z} = 2\pi\begin{pmatrix}
		\rho_{0}-\tilde{\sigma}_{0} & \sigma_{0}-\tilde{\rho}_{0} \\
		\tilde{\sigma}_{0} & \tilde{\rho}_{0}
	\end{pmatrix},
\end{align}
as well as two other matrices
\begin{align}\label{eq:V}
	\mathcal{V} = \begin{pmatrix}
		v_{1} & 0 \\
		0 & v_{2}
	\end{pmatrix}, \quad \mathcal{R} = \begin{pmatrix}
		2\rho_{0} & 0 \\
		0 & 2\sigma_{0}
	\end{pmatrix},
\end{align}
we eventually obtain
\begin{align}
	\label{eqn:Matrix_split_RHS1}
	\begin{pmatrix}
		\frac{\partial \mu_{\!_F}}{\partial Q} & \ \frac{\partial \mu_{\!_F}}{\partial B}  \\
		\frac{\partial \mu_{\!_B}}{\partial Q} &\ \frac{\partial \mu_{\!_B}}{\partial B} 
	\end{pmatrix} = \pi\hbar\left( \mathcal{Z}^{-1}\right)^{T} \mathcal{V} \mathcal{R},
\end{align}
where we have used Eq.~(\ref{eq:Ephixiomega}). We have thus obtained a convenient expression for the first matrix on the right-hand side of Eq.~(\ref{eqn:Matrix_split}).

Now we evaluate the remaining matrix on the right-hand side of Eq.~(\ref{eqn:Matrix_split}). To do so, from Eq.~(\ref{eqn:GS_density}) we first evaluate the derivative
\begin{align}
	\frac{\partial n}{\partial Q} = 2\rho_{0} + \int_{-Q}^{Q}dk\;\!\rho'_{\Q}(k).
\end{align}
Here we have used the parity, $\rho(k)=\rho(-k)$. Moreover, $\rho'_{\Q}(k)=\partial \rho(k,Q,B)/\partial Q$, see the comment below Eq.~(\ref{eq:E0}). Using the results derived in Appendix \ref{appendix:relations}, the integral in the above equation can be simplified as
\begin{align}\label{eq:intint}
\int_{-Q}^{Q}dk\;\!{\rho'_{\Q}(k)} ={}&-2\rho_0\int_{-B}^{B}dk\;\! \sigma(k)\theta_+(k,Q)\notag\\
={}&4\pi\rho_0^2-2\rho_0.
\end{align}
Here and in the following we use the abbreviation 
\begin{align}
\theta_{\pm}(k,Q)=\theta'(2k-2Q)\pm\theta'(2k+2Q).
\end{align}
The second line of Eq.~(\ref{eq:intint}) follows from Eq.~(\ref{eqn:Ground_state_root_1}) taken  at $k=\pm Q$. In a similar way we can obtain the remaining derivatives. They are
\begin{subequations}\label{eqn:density_derivatives}
\begin{gather}
	\frac{\partial n}{\partial Q} = 4\pi\rho_{0}^2
,\quad\frac{\partial n}{\partial B} = 4\pi\sigma_{0}^{2},\\
\frac{\partial n_{\B}}{\partial Q} = 4\pi\rho_{0}\tilde{\sigma}_{0},\quad \frac{\partial n_{\B}}{\partial B} = 4\pi\sigma_{0}\tilde{\rho}_{0}.
\end{gather}
\end{subequations}
Since $n_{\F}=n-n_{\B}$, the matrix of derivatives can be expressed as
\begin{align}\label{eq:Jmatrix}
 \mathcal{J} = \begin{pmatrix}
\frac{\partial n_{\!_F}}{\partial Q} &\  \frac{\partial n_{\!_B}}{\partial Q} \\
\frac{\partial n_{\!_F}}{\partial B} &\  \frac{\partial n_{\!_B}}{\partial B}
\end{pmatrix} =4\pi\begin{pmatrix}
		\rho_{0}(\rho_{0}-\tilde{\sigma}_{0}) &  \rho_{0}\tilde{\sigma}_{0} \\
		\sigma_{0}(\sigma_{0}-\tilde{\rho}_{0}) & \sigma_{0}\tilde{\rho}_{0}
	\end{pmatrix}.
\end{align}
The matrix $ \mathcal{J}$ can be understood as the (transposed) Jacobian matrix of the transformation from $(Q,B)$ to $(n_{\F}, n_{\B})$. We note that $ \mathcal{J} =  \mathcal{R} \mathcal{Z}^{T}$. The matrix of the inverse transformation is given by 
\begin{equation}
	\label{eqn:Matrix_split_RHS2}
	 \mathcal{J}^{-1} = \begin{pmatrix}
		\frac{\partial Q}{\partial n_{\!_F}} & \frac{\partial B}{\partial n_{\!_F}}\\
		\frac{\partial Q}{\partial n_{\!_B}} & \frac{\partial B}{\partial n_{\!_B}}
	\end{pmatrix}.
\end{equation}
The latter is equal to the transpose of the second matrix on the right-hand side of Eq.~(\ref{eqn:Matrix_split}). Thus using Eqs.~(\ref{eqn:Matrix_split_RHS1}) and (\ref{eqn:Matrix_split_RHS2}), we can rewrite Eq.~(\ref{eqn:Matrix_split}) as
\begin{align}
	\mathcal{M} =\left( \mathcal{Z}^{-1}\right)^{T} \mathcal{V} \mathcal{R}( \mathcal{J}^{-1})^T.
\end{align}
Using $ \mathcal{J} =  \mathcal{R} \mathcal{Z}^{T}$, a further simplification occurs, leading to the exact expression 
\begin{align}\label{eqn:Exact_Relation_ChemicalPot}
	\mathcal{M} = ( \mathcal{Z}^{T})^{-1} \mathcal{V} \mathcal{Z}^{-1}.
\end{align}
Explicitly we have
\begin{subequations}\label{eq:Mmatrixelements}
\begin{gather}\mathcal{M}_{\FF} = \frac{4\pi^2}{(\det{ \mathcal{Z}})^{2}}\left(v_{1}\tilde{\rho}_{0}^{2}+v_{2}\tilde{\sigma}_{0}^{2}\right),\\
\mathcal{M}_{\BB}  = \frac{4\pi^2}{(\det{ \mathcal{Z}})^{2}}\left[v_{1}(\tilde{\rho}_{0}-\sigma_{0})^2+v_{2}(\rho_{0}-\tilde{\sigma}_{0})^{2}\right],\\
\mathcal{M}_{\FB}  = \frac{4\pi^2}{(\det{ \mathcal{Z}})^{2}}\left[v_{1}\tilde{\rho}_{0}(\tilde{\rho}_{0}-\sigma_{0})+v_{2}\tilde{\sigma}_{0}(\tilde{\sigma}_{0}-\rho_{0})\right].
\end{gather}
\end{subequations}
Note that the compressibility matrix \cite{Note1} has three independent components since $\mathcal{M}_{\BF} =\mathcal{M}_{\FB}$, as it follows from the definition (\ref{eqn:Matrix_split_RHS1}). The derivation presented in this section can be related to the one for finite-size corrections in the one-dimensional Hubbard model \cite{essler}. 

As a side remark we note that the stability of the system toward the demixing is controlled by the compressibility matrix. If it is positive definite, the Bose-Fermi mixture is stable \cite{viverit_zero-temperature_2000,imambekov_applications_2006}. Equation~(\ref{eqn:Exact_Relation_ChemicalPot}) implies $\det \mathcal{M} >0$ if $v_1 v_2>0$, which is the case. Therefore, $\mathcal{M}$ is positive definite. Previously this was checked only numerically \cite{imambekov_exactly_2006}. Therefore, the integrable case of the Bose-Fermi mixture is stable against demixing, contrary to the mean-field prediction \cite{das_2003,cazalilla_2003}.

\section{Additional relations due to Galilean Invariance}\label{sec:additionalrelations}

Assuming the compressibility matrix is known, i.e., its three independent matrix elements, the decomposition (\ref{eqn:Exact_Relation_ChemicalPot}) leads to three equations that connect the known matrix elements and six quantities, two velocities, $v_1$ and $v_2$, and four elements of the dressed charge matrix, $\rho_0,\sigma_0,\tilde\rho_0$, and $\tilde\sigma_0$. In this section we find additional constraints that follow from Galilean invariance of the model (\ref{eq:Hamiltonian}). 

We begin by differentiating Eqs.~(\ref{eqn:Ground_state_root}) with respect to $Q$, yielding the set of equations
\begin{subequations}
	\label{eqn:BF_GS_qder}
	\begin{align}
		\label{eqn:BF_GS_qder_1}
		\mathcal{F}[\rho'_{\Q},\sigma'_{\Q},B] &= 0, \\
		\label{eqn:BF_GS_qder_2}
		\mathcal{F}[\sigma'_{\Q},\rho'_{\Q},Q] &= -\frac{\rho_{0}}{\pi}\theta_{+}(k,Q).
	\end{align}
\end{subequations}
In Eq.~(\ref{eqn:Ground_state_root}) we should keep in mind that $\rho$ and $\sigma$ depend on three arguments, as discussed below Eq.~(\ref{eq:E0}). In a similar way, differentiating the Eqs.~(\ref{eqn:Ground_state_root}) with respect to $B$, we obtain
\begin{subequations}
	\label{eqn:BF_GS_bder}
	\begin{align}
		\label{eqn:BF_GS_bder_1}
		\mathcal{F}[\rho'_{\B},\sigma'_{\B},B] &= -\frac{\sigma_{0}}{\pi}\theta_{+}(k,B), \\
		\label{eqn:BF_GS_bder_2}
		\mathcal{F}[\sigma'_{\B},\rho'_{\B},Q] &= 0.
	\end{align}
\end{subequations}
Using the linearity of the functional, we can combine Eqs.~(\ref{eqn:BF_GS_qder}) and (\ref{eqn:BF_GS_bder}) and obtain 
\begin{subequations}
	\label{eqn:Comb1}
	\begin{align}
		\mathcal{F}\left[\frac{\rho'_{\B}  + a_1\rho'_{\Q}}{\sigma_{0}},\frac{\sigma'_{\B}+ a_1\sigma'_{\Q}}{\sigma_{0}},B\right] &= -\frac{1}{\pi}\theta_{+}(k,B), \\
		\mathcal{F}\left[\frac{a_2\sigma'_{\B}  + \sigma'_{\Q}}{\rho_{0}},\frac{a_2\rho'_{\B}+ \rho'_{\Q}}{\rho_{0}},Q\right] &= -\frac{1}{\pi}\theta_{+}(k,Q),
	\end{align}
\end{subequations}
where $a_1$ and $a_2$ are arbitrary real numbers. Additional equations arise by  differentiating Eqs.~(\ref{eqn:BF_exc_spec}) with respect to $k$. This gives
\begin{subequations}
	\label{eqn:BF_exc_kder}
	\begin{align}
		\label{eqn:BF_exc_kder_1}
		\mathcal{F}[\xi'_{k},\omega '_{q},B] &= \frac{\hbar^2}{m}+\frac{\omega_{0}}{\pi}\theta_{+}(k,B), \\
		\label{eqn:BF_exc_kder_2}
		\mathcal{F}[\omega '_{k},\xi'_{q},Q] &= \frac{\xi_{0}}{\pi}\theta_{+}(k,Q).
	\end{align}
\end{subequations}
Here 
\begin{align}
\xi_0=\xi(Q),\quad \omega_{0}=\omega(B).
\end{align}
Equations~(\ref{eqn:Comb1}) can be combined with Eqs.~(\ref{eqn:BF_exc_kder}) to obtain
\begin{subequations}\label{eq:combination}
\begin{align}
\mathcal{F}\left[a\frac{\xi'_{k}}{\omega_{0}} + a\frac{\rho'_{\B}  +  a_1\rho'_{\Q}}{\sigma_{0}},a \frac{\omega'_{q}} {\omega_{0}}+a\frac{\sigma'_{\B}+a_1\sigma'_{\Q}}{\sigma_{0}},B\right]
&= \frac{a\hbar^2}{m\omega_{0}}, \\
		\mathcal{F}\left[\frac{\omega '_{k}}{\xi_{0}} + \frac{a_2\sigma'_{\B}  + \sigma'_{\Q}}{\rho_{0}},\frac{\xi'_{q}}{\xi_{0}}+\frac{a_2\rho'_{\B}+ \rho'_{\Q}}{\rho_{0}},Q\right]  &= 0,
	\end{align}    
\end{subequations}
where $a$ is arbitrary. Choosing $a = \frac{\omega_{0}}{\xi_{0}}, a_1 = \frac{\sigma_{0}\xi_{0}}{\rho_{0}\omega_{0}} = \frac{v_{1}}{v_{2}}$, and $a_2 = \frac{v_{2}}{v_{1}}$, Eqs.~(\ref{eq:combination}) up to a multiplicative constant become equivalent to Eqs.~(\ref{eqn:Ground_state_root}). Using the uniqueness of the solutions of Eqs.~(\ref{eqn:Ground_state_root}), see Appendix~\ref{appendix-functional}, we obtain the relations
\begin{subequations}\label{eqn:BF_Der_rel1}
	\begin{align}
		\label{eqn:BF_Der_rel1_1}
		\frac{\xi'_{k}}{2\pi\hbar} + v_{1}\rho'_{\Q}+v_{2}\rho'_{\B}= \frac{\hbar}{m}\rho, \\
		\label{eqn:BF_Der_rel1_2}
		\frac{\omega'_{k}}{2\pi\hbar} + v_{1}\sigma'_{\Q}+v_{2}\sigma'_{\B} = \frac{\hbar}{m}\sigma.
	\end{align} 
\end{subequations}
Here we have used the definition (\ref{eq:velocities}). Similarly, it is possible to obtain                                      
\begin{subequations}\label{eqn:BF_Der_rel2}
\begin{gather}
\label{eqn:BF_Der_rel2_1}
2\pi\hbar\rho'_{k} + \frac{1}{v_{1}}\xi'_{\Q}+\frac{1}{v_{2}}\xi'_{\B} = 0, \\
\label{eqn:BF_Der_rel2_2}
2\pi\hbar\sigma'_{k} + \frac{1}{v_{1}}\omega'_{\Q}+\frac{1}{v_{2}}\omega'_{\B} = 0.
\end{gather}
\end{subequations}

Integrating the relations (\ref{eqn:BF_Der_rel1_1}) and (\ref{eqn:BF_Der_rel1_2}) with respect to $k$, respectively, from $-Q$ to $Q$ and from $-B$ to $B$, we obtain 
\begin{subequations}	\label{eqn:BF_der_good_comb}
\begin{gather}
	\label{eqn:BF_der_good_comb_1}
	v_{1}\frac{\partial n}{\partial Q} + v_{2}\frac{\partial n}{\partial B} = \frac{\hbar n}{m},\\
	v_{1}\frac{\partial n_{\B}}{\partial Q} + v_{2}\frac{\partial n_{\B}}{\partial B} = \frac{\hbar n_{\B}}{m}.
\end{gather}
\end{subequations}
Here we have used the densities expressed in the form (\ref{eqn:GS_density}). Using the relations (\ref{eqn:density_derivatives}), Eqs.~(\ref{eqn:BF_der_good_comb}) eventually become 
\begin{subequations}\label{eqn:Gal_inv_rel}
\begin{gather}\label{eqn:Gal_inv_rel_1}
	v_{1}\rho_{0}^2 + v_{2}\sigma_{0}^2 = \frac{\hbar n}{4\pi m},\\
	\label{eqn:Gal_inv_rel_2}
	v_{1}\rho_{0}\tilde{\sigma}_{0} + v_{2}\sigma_{0}\tilde{\rho}_{0} = \frac{\hbar n_{\B}}{4\pi m}.
\end{gather}
\end{subequations}
Using the definitions (\ref{eq:velocities}) we obtain another form
\begin{subequations}\label{eqn:Gal_inv_relanother}
\begin{gather}\label{eqn:Gal_inv_rel_another1}
	\rho_{0}\xi_0 + \sigma_{0}\omega_0 = \frac{\hbar^2 n}{2 m},\\
	\label{eqn:Gal_inv_rel_another2}
	\tilde{\sigma}_{0}\xi_0 + \tilde{\rho}_{0}\omega_0 = \frac{\hbar^2 n_{\B}}{2m}.
\end{gather}
\end{subequations}
The constraints (\ref{eqn:Gal_inv_rel}) and (\ref{eqn:Gal_inv_relanother}) rely on Galilean invariance of the model (\ref{eq:Hamiltonian}). In this case, the Bethe-ansatz equations for the density of rapidities are characterized by a constant term on the right-hand side of Eq.~(\ref{eqn:Ground_state_root_1}) as well as by a linear term on the right-hand side of Eq.~(\ref{eqn:BF_exc_spec_1}) that is equivalent to a quadratic dispersion in the dressed energy equation (\ref{eqn:BF_dressed_T0_1}). We should contrast the present two-component system to one-component Galilean invariant liquids, with the Lieb--Liniger model as a prototypical example. The relation (\ref{eqn:Gal_inv_rel_1}) reduces in the one-component case to \cite{korepin}
\begin{align}\label{eq:mvK}
v \rho_0^2=\frac{\hbar n}{4\pi m}.
\end{align}
The relation (\ref{eq:mvK}) is equivalent to $mvK=\pi\hbar n$ mentioned in the introduction. Therefore, the well-known constraint (\ref{eq:mvK}) is replaced by two other relations given by Eqs.~(\ref{eqn:Gal_inv_rel})  in the two-component liquids.

The relations (\ref{eqn:Gal_inv_rel}) give two additional constraints on the velocities and the matrix elements of the dressed charge matrix. Combined with the three independent relations that follow from Eq.~(\ref{eqn:Exact_Relation_ChemicalPot}), they give five independent equations for six unknowns. In order to solve this system, we need one more equation which will be obtained from the study of the Drude weight.

\section{The Drude weight matrix}\label{sec:Drudeweight}

In addition to the compressibility, the Drude weight is another thermodynamic quantity that characterizes the system \cite{kohn_theory_1964,shastry_twisted_1990,sutherland_adiabatic_1990}. It can be understood as a measure of the energy change upon imposing the twisted boundary conditions. For our two-component system with bosons and fermions, the twisted boundary conditions for the wave function are given by 
\begin{align}\label{eq:TWBC}
	\Psi(\ldots,x_{j}&+L,\ldots) =\Psi(\ldots,x_{j},\ldots)\notag\\
	&\times\begin{cases}
		\exp(i\varphi_{\B}),\quad 1\le j\le N_{\B},\\
		\exp(i\varphi_{\F}),\quad N_{\B}+1\le j\le N.
	\end{cases}
\end{align}
The resulting change in the ground-state energy will appear as a finite-size effect, i.e., in the subleading order. The Drude weight is a square matrix 
\begin{align}\label{eq:Drudematrix}
	\mathcal{D}=\begin{pmatrix}
		\mathcal{D}_{\FF} & \mathcal{D}_{\FB}\\
		\mathcal{D}_{\BF} & \mathcal{D}_{\BB}
	\end{pmatrix},
\end{align}
with the matrix elements defined by
\begin{align}\label{eq:Drudematrixelements}
	\mathcal{D}_{ss'} = \frac{\pi L}{\hbar}\left.\frac{\partial^2 (E_t-E_0)}{\partial \varphi_{s}\partial \varphi_{s'}}\right|_{\varphi_{s}\to 0,\varphi_{s'}\to 0},
\end{align}
where $s,s' \in \{F,B\}$, $E_t$ is the ground-state energy for the system with twisted boundary conditions, and $E_0$ is the ground-state energy for the system with periodic boundary conditions. We emphasize that the difference $E_t-E_0$ is to be taken for the same particle density of the two systems.

For a Bose-Fermi mixture with $N_{\B}$ bosons and $N_{\F}=N-N_{\B}$ fermions, the ground state equations for the Bethe ansatz are given by Eqs.~(\ref{eqn:DBAElog}) where the quantum numbers assume the values (\ref{eqn:quantumnumbers}). Under the twisted boundary conditions (\ref{eq:TWBC}), the Bethe-ansatz equations are given by 
\begin{subequations}\label{eqn:DBAlogtwisted}
\begin{gather}
	L\tilde{k}_{j} = 2\pi I_{j} +\varphi_{\F} + \sum_{l=1}^{N_{\B}}\theta(2\tilde{k}_{j}-2\tilde{\Lambda}_{l}), \quad \ j = 1,2,\ldots,N,\\
	2\pi J_{l} +\varphi_{\F}-\varphi_{\B} = \sum_{j=1}^{N}\theta(2\tilde{\Lambda}_{l}-2\tilde{k}_j),\quad l = 1,2,\ldots,N_{\B}. 
\end{gather}
\end{subequations}
The actual derivation of Eqs.~(\ref{eqn:DBAlogtwisted}) is a mathematically difficult task \cite{shastry_twisted_1990,sutherland}. In the ground state, the quantum numbers in Eqs.~(\ref{eqn:DBAlogtwisted}) are still given by Eqs.~(\ref{eqn:quantumnumbers}) for small $\varphi_{\F}$ and $\varphi_{\B}$ \footnote{Equations~(\ref{eqn:DBAlogtwisted}) in the limiting cases give expected results. For $N=N_{\B}$, our model reduces to the Lieb--Liniger one subject to the boundary condition (\ref{eq:TWBC}). The quasimomenta of the latter $\tilde k_j$ are obtained from the equations
\begin{align}\label{eq:DBA-log-LL-twist}
L \tilde k_j=2\pi I_j+\varphi_{\B}+\sum_{l=1}^{N}\theta(\tilde k_j-\tilde k_l).
\end{align}
We have verified that the quasimomenta of the system (\ref{eq:DBA-log-LL-twist}) for a given $\varphi_{\B}$ coincide with the ones obtained after solving Eqs.~(\ref{eqn:DBAlogtwisted}) for the same $\varphi_{\B}$, at $N=N_{\B}$ and arbitrary $\varphi_F$. Here we consider small twist angles that do not perturb the ground-state quantum numbers $I_j$ and $J_l$ given by Eq.~(\ref{eqn:quantumnumbers}). On the other hand, in the case $N_{\B}=0$, Eqs.~(\ref{eqn:DBAlogtwisted}) describe free fermions that obey the twisted boundary condition (\ref{eq:TWBC}) as long as the untwisted case $\varphi_{\F}=0$ obeys the periodic boundary conditions. This is the case for odd $N_{\F}$ as we discussed below Eq.~(\ref{eqn:quantumnumbers}).}.

The system with twisted boundary conditions has a conserved, nonzero momentum, in the ground state. It is given by 
\begin{align}\label{eq:Pt}
P_t=\hbar\sum_{j=1}^{N}\tilde k_j=\hbar(n_{\F}\varphi_{\F}+ n_{\B}\varphi_{\B}),
\end{align}
as it directly follows from Eqs.~(\ref{eqn:DBAlogtwisted}). The evaluation of the  ground-state energy begins by subtracting the ground-state equations (\ref{eqn:DBAElog}) without the twist from those of Eqs.~(\ref{eqn:DBAlogtwisted}) with the twist. We obtain
\begin{subequations}\label{eq:J1J2}
	\begin{align}
L\Delta k_{j} = \varphi_{\F} + \sum_{l=1}^{N_{\B}}\theta'(2k_{j}-2\Lambda_{l})(2\Delta k_{j} - 2\Delta \Lambda_{l}),\\
		\varphi_{\F}-\varphi_{\B} = \sum_{j=1}^{N}\theta'(2\Lambda_{l}-2k_{j})(2\Delta \Lambda_{l}-2\Delta k_{j}). 
	\end{align}
\end{subequations}
Here we have used $\Delta k_j=\tilde k_j-k_j$ and $\Delta\Lambda_l=\tilde\Lambda_l-\Lambda_l$, and in the right-hand side we neglected higher-order terms in $\Delta k_j$ and $\Delta\Lambda_l$. Let us assume that the Fermi rapidities $Q$ and $B$ shift under the twisted boundary conditions to $Q\to Q+ \delta_{1}$ and $B \to B+\delta_{2}$. Defining the shift functions $J_{1}(k_{j}) = L\rho(k_{j})\Delta k_{j}$ and $J_{2}(\Lambda_{l}) = L\sigma(\Lambda_{l})\Delta \Lambda_{l}$, Eqs.~(\ref{eq:J1J2}) in the thermodynamic limit become 
\begin{subequations}\label{eq:J1J1int}
	\begin{gather}
		J_{1}(k) + \frac{1}{\pi}\int_{-B}^{B}d\Lambda\;\! \theta'(2k-2\Lambda)J_{2}(\Lambda) = \frac{\varphi_{\F}}{2\pi}, \\
		J_{2}(k)+ \frac{1}{\pi}\int_{-Q}^{Q}dk \ \theta'(2\Lambda-2k)J_{1}(k) = \frac{\varphi_{\B}-\varphi_{\F}}{2\pi}.
	\end{gather}
\end{subequations}
It is simple to verify that the solution of Eqs.~(\ref{eq:J1J1int}) can be expressed in terms of linear combinations of previously introduced functions, 
\begin{subequations}\label{eq:shiftfunctions}
	\begin{gather}
		J_{1}(k) = (\varphi_{\B}-\varphi_{\F})\tilde{\sigma}(k) +\varphi_{\F}\rho(k), \\
		J_{2}(k) = (\varphi_{\B}-\varphi_{\F})\tilde{\rho}(k) +\varphi_{\F}\sigma(k). 
	\end{gather}
\end{subequations}
Here $\rho(k)$, $\sigma(k)$, $\tilde\rho(k)$, and $\tilde\sigma(k)$ are defined by Eqs.~(\ref{eq:BEcompact}) [or equivalently Eqs.~(\ref{eqn:Ground_state_root})] and Eqs.~(\ref{eqn:BF_dual_root}).

At the Fermi rapidities, Eqs.~(\ref{eq:shiftfunctions}) satisfy
\begin{subequations}
	\label{eqn:delta_value}
	\begin{gather}
		J_{1}(Q) = (\varphi_{\B}-\varphi_{\F})\tilde{\sigma}(Q) +\varphi_{\F}\rho(Q) = L\rho(Q)\delta_{1},\\
		J_{2}(B) = (\varphi_{\B}-\varphi_{\F})\tilde{\rho}(B) +\varphi_{\F}\sigma(B) = L\sigma(B)\delta_{2}, 
	\end{gather}
\end{subequations}
where the last equalities in both equations follow from the definition of $J_{1}(k)$ and $J_{2}(k)$. They thus express $\delta_1$ and $\delta_{2}$ in terms of the twist phases and $\rho(Q)$, $\sigma(B)$, $\tilde\rho(B)$, and $\tilde\sigma(Q)$. We observe that $\delta_1$ and $\delta_{2}$ scale as $1/L$, which means that their contribution will appear as a finite size effect. The shift functions (\ref{eq:shiftfunctions}) also show that the negative Fermi rapidities shift in the same direction as positive ones and by the same amounts, $-Q \to -Q+\delta_{1}$ and $-B \to -B+\delta_{2}$, since all $\rho(k)$, $\sigma(k)$, $\tilde\rho(k)$, and $\tilde\sigma(k)$ are even functions. We can thus write the ground-state Bethe-ansatz equations for the twisted boundary conditions in the form
\begin{subequations}\label{eq:rtst}
\begin{gather}
\rho_{t}(k) + \frac{1}{\pi}\int_{-B+\delta_{2}}^{B+\delta_{2}}dq\;\! \theta'(2k-2q) \sigma_{t}(q) = \frac{1}{2\pi},\\
\sigma_{t}(k) + \frac{1}{\pi}\int_{-Q+\delta_{1}}^{Q+\delta_{1}}dq\;\!\theta'(2k-2q) \rho_{t}(q) = 0.
	\end{gather}
\end{subequations}
The corresponding ground-state energy is given by 
\begin{align}\label{eq:Et}
E_t= \frac{\hbar^2 L}{2m}\int_{-Q+\delta_{1}}^{Q+\delta_{1}}dk\;\! k^2 \rho_{t}(k).
\end{align}
It depends on the twisting angles $\varphi_{\F}$ and $\varphi_{\B}$. If they are zero, $E_t$ coincides with the ground-state energy $E_0$ of the system with  periodic boundary conditions.

On general grounds $E_t$ should be an analytic function of $\varphi_{\F}$ and $\varphi_{\B}$. Moreover, its leading contribution should arise at second (or higher) orders in the twisting angles. Indeed, Eqs.~(\ref{eqn:DBAlogtwisted}) are invariant to the transformation
\begin{align}
\varphi_{\F}\to -\varphi_{\F},\quad \varphi_{\B}\to -\varphi_{\B},\quad \tilde{k}_j\to-\tilde{k}_j,\quad \tilde{\Lambda}_j\to-\tilde{\Lambda}_j,
\end{align}
as the quantum numbers are distributed symmetrically around zero. Therefore, the ground-state energy $E_t=\frac{\hbar^2}{2m}\sum_{j=1}^{N}\tilde{k}_j^2$ is an even function of $\varphi_{\F}$ and $\varphi_{\B}$. This is physically plausible as it is not expected that the energy changes if the signs of both twisting angles are reversed simultaneously. The resulting Taylor series has the quadratic form
\begin{align}
		E_{t} =E_0+  \frac{1}{2}\begin{pmatrix}
			\varphi_{\F} & \varphi_{\B}
		\end{pmatrix}\begin{pmatrix}
			\frac{\partial^2 E_{t}}{\partial \varphi_{\F}^2}  \frac{\partial^2 E_{t}}{\partial \varphi_{\F}\varphi_{\B}}\\
			\frac{\partial^2 E_{t}}{\partial \varphi_{\B}\varphi_{\F}}  \frac{\partial^2 E_{t}}{\partial \varphi_{\B}^2}\\
		\end{pmatrix}\begin{pmatrix}
			\varphi_{\F}\\
			\varphi_{\B}
		\end{pmatrix}+\ldots.
\end{align}
The actual evaluation of Eq.~(\ref{eq:Et}) at small $\delta_1$ and $\delta_2$ is involved and presented in Appendixes \ref{sec:drude_weight_proof2} and \ref{sec:drude_weight_proof}. The final result can be expressed in the compact form
\begin{align}
E_{t}= E_0+\frac{\hbar}{2\pi L}\begin{pmatrix}
			\varphi_{\F} & \varphi_{\B}
		\end{pmatrix} \mathcal{Z} \mathcal{V} \mathcal{Z}^{T}\begin{pmatrix}
			\varphi_{\F} \\
			\varphi_{\B}
		\end{pmatrix}+\ldots,
\end{align}
where $ \mathcal{Z}$ and $ \mathcal{V}$ are given by Eqs.~(\ref{eq:Z}) and (\ref{eq:V}), and the ellipsis denotes higher-order corrections in powers of $1/L$. Therefore, the Drude weight matrix (\ref{eq:Drudematrix}) is
\begin{equation}
	\label{eqn:Exact_Relation_Drude}
	\mathcal{D} =  \mathcal{Z} \mathcal{V} \mathcal{Z}^{T}.
\end{equation} 
Its components are explicitly given by
\begin{subequations}\label{eq:Dwme}
\begin{gather}
		\mathcal{D}_{\FF}=4\pi^2[v_{1}(\rho_{0}-\tilde{\sigma}_{0})^2 + v_{2}(\sigma_{0}-\tilde{\rho}_{0})^2],\\
		\label{eq:Dbb}
		\mathcal{D}_{\BB}= 4\pi^2(v_{1}\tilde{\sigma}_{0}^2 + v_{2}\tilde{\rho}_{0}^2),\\
		\mathcal{D}_{\FB}= 4\pi^2[v_{1}\tilde{\sigma}_{0}(\rho_{0}-\tilde{\sigma}_{0}) + v_{2}\tilde{\rho}_{0}(\sigma_{0}-\tilde{\rho}_{0})].
\end{gather}
\end{subequations}
From the definition (\ref{eq:Drudematrixelements}), we have $\mathcal{D}_{\BF}=\mathcal{D}_{\FB}$. The matrix elements (\ref{eq:Dwme}) are expressed only in terms of the velocities of excitations and four elements of the dressed charge matrix, $\rho_0,\sigma_0,\tilde\rho_0$, and $\tilde\sigma_0$.

The Drude weight matrix (\ref{eq:Drudematrix}) is real and symmetric. It has a positive determinant. Due to Galilean invariance, its matrix elements obey additional sum rules. Indeed, using the relations (\ref{eqn:Gal_inv_rel}), we obtain that Eqs.~(\ref{eq:Dwme}) obey the constraints
\begin{subequations}\label{eq:Dsumrules}
\begin{gather}\label{eq:Dsumrules1}
\mathcal{D}_{\FF} + \mathcal{D}_{\FB} = \frac{\pi \hbar n_{\F}}{m},\\
\label{eq:Dsumrules2}
\mathcal{D}_{\BB} + \mathcal{D}_{\FB}= \frac{\pi \hbar n_{\B}}{m}.
\end{gather}
\end{subequations}
Therefore, the Drude weight matrix has only one independent matrix element, while the others are fixed by the sum rules and Hermitian symmetry. We also note the relation
\begin{align}\label{eq:Dsumrules3}
\mathcal{D}_{\FF} - \mathcal{D}_{\BB} = \frac{\pi \hbar (n_{\F}-n_{\B})}{m}.
\end{align}
The latter is, however, not an independent relation. Any two equations out of Eqs.~(\ref{eq:Dsumrules1}), (\ref{eq:Dsumrules2}), and (\ref{eq:Dsumrules3}) are linearly independent. Interestingly, the relations (\ref{eq:Dsumrules}) agree with those of Ref.~\cite{orignac_2010}, which were obtained in a phenomenological study. Here they were shown in a fully microscopic exact approach. Note that in the one-component Galilean-invariant liquids, the Drude weight is $\pi\hbar n/m$ \cite{sutherland_adiabatic_1990}, consistent with Eqs.~(\ref{eq:Dsumrules}).

We eventually note that the velocities of excitations in Eqs.~(\ref{eq:Dwme}) can be eliminated using the sum rules (\ref{eqn:Gal_inv_rel}). This leads to
\begin{align}
\mathcal{D}_{\BB}=\frac{\pi\hbar n}{m}\left(\frac{n_{\B}}{n} \frac{\rho_0\tilde\rho_0+\sigma_0\tilde\sigma_0}{\rho_0\sigma_0} - \frac{\tilde\rho_0\tilde\sigma_0}{\rho_0\sigma_0}\right),
\end{align}
while the other two components of the Drude weight matrix then follow from Eqs.~(\ref{eq:Dsumrules3}) and (\ref{eq:Dsumrules2}). Therefore, the nontrivial information in the Drude weight matrix is contained in the matrix elements of the dressed change matrix $\mathcal{Z}$, i.e., the solutions of Eqs.~(\ref{eq:BEcompact}) and (\ref{eqn:BF_dual_root}) evaluated at the Fermi rapidities. The same conclusion also holds for the velocities and the matrix elements of the compressibility matrix. Indeed, Eqs.~(\ref{eqn:Gal_inv_rel}) enable us to express the velocities as
\begin{align}\label{eq:lastv1v2}
\begin{pmatrix}
v_1\\
v_2
\end{pmatrix}=\frac{\hbar}{m}(\mathcal{J}^T)^{-1} \begin{pmatrix}
n_{\F}\\
n_{\B}
\end{pmatrix},
\end{align}
where the matrix $\mathcal{J}$ is given by Eq.~(\ref{eq:Jmatrix}). Equation (\ref{eq:lastv1v2}) can be then used in Eqs.~(\ref{eq:Mmatrixelements}) to eliminate the velocities. In the following we discuss another expression for the velocities.

\section{Velocities of excitations}\label{sec:velocities}

If the compressibility and the Drude weight matrices are known, we can express the velocities of excitations in terms of their matrix elements. Indeed, three equations (\ref{eq:Mmatrixelements}) for the compressibility matrix, two conditions (\ref{eqn:Gal_inv_rel}) for Galilean invariance, and one independent matrix element of the Drude matrix, for example Eq.~(\ref{eq:Dbb}), make a system of six independent linear equations that contain six unknowns, $v_1$, $v_2$, $\rho_0$, $\sigma_0$, $\tilde\rho_0$, and $\tilde\sigma_0$. It can be solved for the unknowns which can be expressed in terms of $\mathcal{M}_{\FF}$, $\mathcal{M}_{\BB}$, $\mathcal{M}_{\FB}$, $\mathcal{D}_{\BB}$, $n_{\F}$, and $n_{\B}$. 

Alternatively, the velocities of excitations follow from Eqs.~(\ref{eqn:Exact_Relation_ChemicalPot}) and (\ref{eqn:Exact_Relation_Drude}). They directly lead to the relations
\begin{subequations}\label{eq:detTr}
\begin{gather}
\det (\mathcal{DM})= v_1^2 v_2^2,\\
\mathrm{Tr} (\mathcal{DM})=v_1^2+v_2^2.
\end{gather} 
\end{subequations}
Therefore, once the determinant and the trace of the matrix product $\mathcal{DM}$ are known, the velocities of excitations can be calculated from the system (\ref{eq:detTr}). Equations (\ref{eq:detTr}) also mean that $v_1^2$ and $v_2^2$ are the eigenvalues of the matrix $\mathcal{DM}$. Note that Eqs.~(\ref{eq:detTr}) can also be understood as Vi\`{e}te's formulas of the biquadratic equation
\begin{align}\label{eq:biquadratic}
(v^2)^2 - \mathrm{Tr} (\mathcal{DM}) v^2+\det (\mathcal{DM})=0.
\end{align}
In this formulation, the squares of velocities, $v_1^2$ and $v_2^2$ are the  solutions of Eq.~(\ref{eq:biquadratic}). They are given by
\begin{align}\label{eq:v1v2exact}
(v_{1,2})^2= \frac{\mathrm{Tr} (\mathcal{DM})}{2}\pm \sqrt{\left[\frac{\mathrm{Tr} (\mathcal{DM})}{2}\right]^2-\det (\mathcal{DM})}.
\end{align}
Note that the discriminant is always nonnegative and the velocities are real. Equation (\ref{eq:biquadratic}) can also be understood as the characteristic polynomials of the matrix $\mathcal{DM}$. Therefore, $v_1^2$ and $v_2^2$ nullify the characteristic polynomial of the matrix $\mathcal{DM}$. As a side result, we note that the matrix $\mathcal{DM}$ also satisfies its own characteristic equation, according to the Cayley–Hamilton theorem. This gives the equation
\begin{align}
	(\mathcal{DM})^2-(v_1^2+v_2^2) \mathcal{DM}+v_1^2 v_2^2 \begin{pmatrix}
		1 &0\\
		0 &1
	\end{pmatrix}
	=0.
\end{align}

Equations (\ref{eq:detTr}) can be related to seemingly different relation (\ref{eq:v2mu(n)}) that explicitly contains the density of particles and the compressibility. Namely, for one-component Galilean-invariant liquids, the Drude weight is  $\mathcal{D}=\pi\hbar n/m$~\cite{sutherland_adiabatic_1990}, and Eq.~(\ref{eq:v2mu(n)}) can be expressed as
\begin{align}\label{eq:v2dm}
v^2=\mathcal{D}\mathcal{M}.
\end{align}
Equation (\ref{eq:v2dm}) thus has a form that can be understood as a special case of Eqs.~(\ref{eq:detTr}).

The velocities of excitations in a Bose-Fermi mixture were previously studied using the method of bosonization \cite{cazalilla_2003,orignac_2010}. For the model (\ref{eq:Hamiltonian}), the obtained results should agree with ours. The resulting low-energy Hamiltonian of Ref.~\cite{cazalilla_2003} consists of a sum of two quadratic Hamiltonians for the two isolated subsystems of interacting bosons and fermions that are coupled by the density-density interaction between the two subsystems. The velocities obtained in Ref.~\cite{cazalilla_2003} are consistent with our exact result (\ref{eq:v1v2exact}) only if we set $\mathcal{D}_{\FB}=0$ in Eq.~(\ref{eq:v1v2exact}). The latter approximation is justified only in the case of weak repulsions when  $\mathcal{D}_{\FB}$ is indeed small \cite{chandak_unpub}. In the study \cite{orignac_2010}, on the other hand, the most general phenomenological quadratic Hamiltonian was used. In addition to the described model of Ref.~\cite{cazalilla_2003}, it contains a term that couples the momentum operators of the two subsystems with a prefactor proportional to $\mathcal{D}_{\FB}$. The obtained velocities of Ref.~\cite{orignac_2010} are in agreement with our result (\ref{eq:v1v2exact}). Our study thus indirectly points out that the effective phenomenological low-energy Hamiltonian of the Bose-Fermi mixture of Ref.~\cite{orignac_2010} is consistent with our exact microscopic results. Unlike the effective low-energy Hamiltonian, the microscopic model, for example Eq.~(\ref{eq:Hamiltonian}), only contains density-density type interactions. An  interesting open problem is the derivation of the effective low-energy Hamiltonian starting from the model (\ref{eq:Hamiltonian}) that describes the physics at all energies. Such a study should further strengthen the important role of the off-diagonal Drude-weight matrix element $\mathcal{D}_{\FB}$.

\section{Summary}\label{sec:discussions}

In summary, using the Bethe ansatz we have studied the one-dimensional Bose-Fermi mixture with contact repulsions described by the Hamiltonian (\ref{eq:Hamiltonian}) and obtained several new exact results. The main ones are (i) an exact microscopic derivation of the compressibility \cite{Note1} and Drude weight  matrices, $\mathcal{M}$ and $\mathcal{D}$, see Eqs.~(\ref{eqn:Exact_Relation_ChemicalPot}) and (\ref{eqn:Exact_Relation_Drude}), respectively, (ii) the exact sum rules (\ref{eq:Dsumrules}) and (\ref{eq:Dsumrules3}) for the Drude weights that follow from Galilean invariance, and (iii) the exact relations (\ref{eq:detTr}) that the squared excitation velocities are the eigenvalues of the matrix product $\mathcal{DM}$. Moreover, we showed that both matrices $\mathcal{M}$ and $\mathcal{D}$ can only be expressed in terms of the matrix elements of the dressed charge matrix (\ref{eq:Z}).

An interesting consequence of our results is that once the determinant and the trace of the product $\mathcal{DM} = \mathcal{Z V}^2\mathcal{Z}^{-1}$ are taken, the dressed charge matrix disappears and one obtains a closed form expression for the two excitation velocities, see Eqs.~(\ref{eq:detTr}). The latter is the generalization of the well-known result (\ref{eq:v2mu(n)}) for the sound velocity of one-component one-dimensional quantum liquids \cite{lieb_exact_1963b,haldane_effective_1981}, which was shown exactly for the Bethe-ansatz solvable cases \cite{haldane_demonstration_1981}. The present study can be understood as a nontrivial extension of the work of Haldane \cite{haldane_demonstration_1981} to the two-component Galilean-invariant case.

Here we have studied the two-component system with spin-polarized fermions. It would be interesting to generalize the present study to the mixture of bosons with spin-$\frac{1}{2}$ fermions or to the case of spin-$\frac{1}{2}$ fermions, both of which admit Bethe-ansatz solutions  \cite{lai_yang_exact}. In the latter cases the system of coupled Bethe-ansatz equations are characterized by two kernels unlike one in the present case and one should find a way to deal with it. It would thus be interesting to verify if the eigenvalues of the product of the compressibility and the Drude weight matrices  correspond to the squares of the excitation velocities beyond the Bose-Fermi mixture, and moreover, in nonintegrable models.

\textit{Note added.} Two months after the submission of our manuscript to the public repository arxiv, a related preprint appeared \cite{liu_universal_2026}. There, our results for the Drude weights have been rederived using the thermodynamic Bethe ansatz and generalized hydrodynamics, together with a relation equivalent to our result~(\ref{eq:detTr}).

\section*{Acknowledgments}

This project was supported in part by the Program QuanTEdu-France ANR-22-CMAS-0001 France 2030.
\appendix

\section{Reduction to the Lieb--Liniger model}\label{sec:reductiontoLL}

Apart from the Bose-Fermi mixture, the Hamiltonian (\ref{eq:Hamiltonian}) also describes the Lieb--Liniger model. In this case the wave function only depends on the coordinates of bosons and obeys the bosonic symmetry to the exchange of coordinates. Mathematically speaking, the discrete Bethe-ansatz equations (\ref{eqn:DBAElog}) in the case of no fermions, $N=N_{\B}$, should thus give the quasimomenta $k_j$ that are identical to those of the Lieb--Liniger model. The latter are the solutions of the Bethe-ansatz equations
\begin{align}\label{eq:DBA-LL}
L k_j=2\pi I_j+\sum_{l=1}^{N}\theta(k_j-k_l),\quad j=1,2,\ldots,N,
\end{align}
where the phase shift is given by the expression (\ref{eq:phaseshift}). It is however not at all obvious that, e.g., the ground-state solution of the system~(\ref{eq:DBA-LL}) coincides with the corresponding solution of the system (\ref{eqn:DBAElog}), which in addition contains auxiliary $\Lambda$ quasimomenta. We have verified that the solution of the system (\ref{eq:DBA-LL}) also satisfies the system (\ref{eqn:DBAElog}) when supplemented by certain $\Lambda$'s. As a side result, we obtained the exact relation
\begin{align}
\sum_{j=1}^{N}(\Lambda_j^2-k_j^2)=\frac{N(N-1)}{4}c^2.
\end{align}
Since $\sum_{j=1}^{N}k_j^2\propto N^3$, at the leading order in $N$, $\sum_{j=1}^{N}\Lambda_j^2$ and $\sum_{j=1}^{N}k_j^2$ are identical.

The reduction to the Lieb--Liniger model is more obvious on the level of Eqs.~(\ref{eqn:Ground_state_root}) that apply in the thermodynamic limit. Substituting $\sigma(k)$ from Eq.~(\ref{eqn:Ground_state_root_2}) into Eq.~(\ref{eqn:Ground_state_root_1}), in the limit $B\to\infty$, we obtain
\begin{align}\label{eq:LLlieb}
\rho(k)+\frac{1}{2\pi}\int_{-Q}^{Q}dq\;\!\theta'(k-q)\rho(q)=\frac{1}{2\pi},
\end{align}
which is the expression for the density of quasimomenta in the ground state of the Lieb--Liniger model \cite{lieb_exact_1963a}. The kernel of Eq.~(\ref{eq:LLlieb}) follows from the identity
\begin{align}\label{eq:A4}
\frac{1}{\pi^2}\int_{-\infty}^{\infty} dp\;\! \theta'(2k-2p)\theta'(2p-2q)=-\frac{1}{2\pi}\theta'(k-q).
\end{align}
Here we use Eq.~(\ref{eq:kernel}).

\section{Properties of the operator $\mathcal{F}$}\label{appendix-functional}

Here we discuss some properties of the operator $\mathcal{F}$ introduced by Eq.~(\ref{eq:Fdefinition}). 

(i) The operator is simultaneously linear for both functions. This means, if 
\begin{align}
\mathcal{F}[\rho_{1},\sigma_{1},B] = f(k)\quad\textrm{and}\quad \mathcal{F}[\rho_{2},\sigma_{2},B] = g(k)
\end{align}
then
\begin{equation}
\mathcal{F}[a\rho_{1}+b\rho_{2},a\sigma_{1}+b\sigma_{2},B] = af(k)+bg(k),
\end{equation}
where $a$ and $b$ are real numbers. The proof is obvious.

(ii) Consider the integral equation
\begin{equation}\label{eqn:iterated0}
f(k,Q,B) - \int_{-Q}^{Q}dq\;\!\theta_{2}(k,q)f(q,Q,B) = 0,
\end{equation}
where the (nonzero) kernel $\theta_{2}(k,q)$ is given by 
\begin{align}\label{eq:kernel2}
\theta_{2}(k,q) = \frac{1}{\pi^2}\int_{-B}^{B}dk_{1}\;\!\theta'(2k-2k_{1})\theta'(2k_{1}-2q).
\end{align}
Let us assume that Eq.~(\ref{eqn:iterated0}) only has a trivial solution $f(k,Q,B) = 0$. Then for the two sets of integral equations
\begin{equation}\label{eq1}
\mathcal{F}[f_{1},g_{1},B] = h_{1}(k),\quad \mathcal{F}[g_{1},f_{1},Q] = 0, 
\end{equation}
and
\begin{equation}\label{eq2}
\mathcal{F}[f_{2},g_{2},B] = h_{2}(k),\quad \mathcal{F}[g_{2},f_{2},Q] = 0, 
\end{equation}
we have $f_{1}(k,Q,B)=f_{2}(k,Q,B)$ and $g_{1}(k,Q,B)=g_{2}(k,Q,B)$ if and only if $h_{1}(k)=h_{2}(k)$. The proof of one side of the implication is trivial. The proof of the other side we begin by subtracting Eq.~(\ref{eq2}) from Eq.~(\ref{eq1}), leading to
\begin{align}
\mathcal{F}[f_1-f_2,g_1-g_2,B]=0, \quad \mathcal{F}[g_1-g_2,f_1-f_2,Q]=0.
\end{align}
Expressing $g_1-g_2$ from the second equation and substituting in the first we obtain Eq.~(\ref{eqn:iterated0}) with $f=f_1-f_2$. The latter by the assumption only has a trivial solution, i.e., $f_1=f_2$. Then we also have $g_1=g_2$.

(iii) For even $h_1(k)=h_1(-k)$, the solutions $f_1(k,Q,B)$ and $g_1(k,Q,B)$ of the system (\ref{eq1}) are even functions with respect to $k$. Similarly, for odd $h_1(k)=-h_1(-k)$, the solutions $f_1(k,Q,B)$ and $g_1(k,Q,B)$ of the system (\ref{eq1}) are odd functions with respect to $k$. In the paper we thus have, e.g., $\rho(k,Q,B)$, $\sigma(k,Q,B)$, $\mathcal{E}(k,Q,B)$, and $\varphi(k,Q,B)$ as even functions of $k$, while $\xi(k,Q,B)$ and $\omega(k,Q,B)$ are odd functions of $k$. Note that the derivative with respect to $k$ of a function changes its  parity.

(iv) The listed properties (i), (ii), and (iii) do not rely on a specific choice of the kernel. Let us show that for the kernel (\ref{eq:kernel}), the assumption that the integral equation (\ref{eqn:iterated0}) only has a trivial solution indeed holds. Consider the eigenvalue problem 
\begin{align}\label{eq:ev}
f(k)= \int_{-Q}^{Q}dq\;\!\theta_{2}(k,q)f(q),
\end{align}
where we suppress unnecessary arguments from $f(k)$. Since the integrand in Eq.~(\ref{eq:kernel2}) is positive, we have the inequalities
\begin{align}
0<\theta_2(k,q)< \tilde\theta_{2}(k,q,c)=\frac{c}{\pi}\frac{1}{c^2+(k-q)^2},
\end{align}
which can be obtained by extending the domain of integration to be over the whole real axis, see Eq.~(\ref{eq:A4}). For $-Q\le k\le Q$ we then have a series of inequalities
\begin{align}
|f(k)|\le{}& \int_{-Q}^{Q}dq\;\!\theta_{2}(k,q)|f(q)| \le \int_{-Q}^{Q}dq\;\!\tilde\theta_{2}(k,q,c)|f(q)|\notag\\
\le{}&\int_{-Q}^{Q}dq\int_{-Q}^{Q}dq_1\tilde\theta_2(k,q,c) \tilde\theta_2(q,q_1,c)|f(q_1)|\notag\\
\le{}& \int_{-Q}^{Q}dq_1\int_{-\infty}^{\infty}dq\;\!\tilde\theta_2(k,q,c) \tilde\theta_2(q,q_1,c)|f(q_1)| \notag\\
={}&\int_{-Q}^{Q}dq_1\;\!\tilde\theta_2(k,q_1,2c)|f(q_1)|.
\end{align}
Repeating this procedure $j$ times, we obtain
\begin{align}
|f(k)|\le{}&\int_{-Q}^{Q}dq_1\;\!\tilde\theta_2(k,q_1,2^j c)|f(q_1)|\notag\\
\le{}&\frac{1}{2^j \pi c}\int_{-Q}^{Q}dq_1\;\!|f(q_1)|.
\end{align}
For $j\to\infty$, the right-hand side tends to zero (for normalizable $f(k)$ that we need). Therefore, the only possibility to have the inequality satisfied is $f(k)=0$. This implies that the eigenvalue problem (\ref{eq:ev}), i.e., the one of Eq.~(\ref{eqn:iterated0}), only has a trivial solution $f=0$ for the kernel (\ref{eq:kernel}). The Fredholm alternative theorem \cite{book-integralequations} then guarantees that the inhomogeneous equation
\begin{align}
f(k,Q,B) - \int_{-Q}^{Q}dq\;\!\theta_{2}(k,q)f(q,Q,B)=h(k)
\end{align}
has a unique solution for any $h(k)\neq 0$ continuous on $[-Q,Q]$. Applied to our case, this means that the functions $\rho(k)$, $\sigma(k)$, $\tilde\rho(k)$, $\tilde\sigma(k)$, $\xi(k)$, $\omega(k)$, $\mathcal{E}(k)$, $\varphi(k)$, etc.~are the unique (and nonzero) solutions of the corresponding integral equations.

(v) Here we show that the solutions of Eqs.~(\ref{eqn:BF_exc_spec}) satisfy $\xi(Q)>0$ and $\omega(B)>0$, where $Q>0$ and $B>0$. We recall that $\xi(k)$ and $\omega(k)$ are odd functions. First we note that the combination of Eqs.~(\ref{eqn:BF_exc_spec_1}) and (\ref{eqn:BF_exc_spec_2}) gives
\begin{align}\label{eq:xiit}
\xi(k)=\frac{\hbar^2 k}{m}+\int_{0}^{Q}dq\;\![\theta_2(k,q)-\theta_2(k,-q)]\xi(q),
\end{align}
where $\theta_2$ is defined by Eq.~(\ref{eq:kernel2}). The kernel of Eq.~(\ref{eq:xiit}) is positive for $k>0$ with $\theta'$ given by Eq.~(\ref{eq:kernel}). Indeed,
\begin{align}\label{eq:kernel3}
\theta_2(k,q)-\theta_2(k,-q)={}&\frac{1}{\pi^2}\int_{0}^{B} dk_1\;\! \vartheta(k,q,k_1),
\end{align}
where
\begin{align}
\vartheta(k,q,k_1)={}&[\theta'(2k+2k_1)-\theta'(2k-2k_1)]\notag\\
&\times [\theta'(2k_1+2q)-\theta'(2k_1-2q)].
\end{align} 
For $k> 0$, $q> 0$, and $k_1>0$, we have $\vartheta(k,q,k_1)>0$ and thus the kernel (\ref{eq:kernel3}) is positive. Therefore, the Neumann series of the integral equation (\ref{eq:xiit}) is a sum of strictly positive terms for $k>0$, which implies $\xi(k)>\hbar^2 k/m$ at $k>0$. The other inequality follows from Eq.~(\ref{eqn:BF_exc_spec_2}),
\begin{align}\label{eq:omegaineq}
\omega(B)=\frac{1}{\pi}\int_{0}^{Q}dq\;\![\theta'(2B+2q)-\theta'(2B-2q)]\xi(q)>0,
\end{align}  
since both, the kernel and $\xi(q)$ are positive for $q>0$. We note that a stronger inequality holds that can be obtained by substituting the inequality for $\xi(k)$ in Eq.~(\ref{eq:omegaineq}). We thus showed 
\begin{subequations}
\begin{align}\label{eq:xiineq}
\xi(Q)>{}&\frac{\hbar^2 Q}{m}>0,\\
\omega(B)>{}&\frac{\hbar^2 B}{\pi m}\biggl[\arctan\left(\frac{2Q+2B}{c}\right)+\arctan\left(\frac{2Q-2B}{c}\right) \notag\\ &-\frac{c}{4B}\ln\left(\frac{c^2+4(Q+B)^2}{c^2+4(Q-B)^2}\right)\biggr]>0.
\end{align}
\end{subequations}

A similar procedure can be used to show that the solutions of Eqs.~(\ref{eqn:Ground_state_root}) satisfy
\begin{subequations}
\begin{align}
\rho(Q)>{}&\frac{1}{2\pi},\\ \sigma(B)>{}&\frac{1}{2\pi^2}\!\left[\arctan\left(\frac{2Q+2B}{c}\right) + \arctan\left(\frac{2Q-2B}{c}\right)\right]\notag\\
>{}&0.
\end{align}
\end{subequations}
These inequalities imply that the velocities (\ref{eq:velocities}) satisfy $v_1>0$ and $v_2>0$ at finite $c$, $B$, and $Q$.

\section{Some properties of integral equations}\label{appendix:relations}

Consider a set of coupled equations given by 
\begin{subequations}
	\begin{gather}
		\rho_{1}(k) + \frac{1}{\pi}\int_{-B}^{B}dq\;\!\theta'(2k-2q) \sigma_{1}(q)= g_{1}(k),\\
		\sigma_{1}(k) + \frac{1}{\pi}\int_{-Q}^{Q}dq\;\!\theta'(2k-2q) \rho_{1}(q)= 0,
	\end{gather}
\end{subequations}
where the kernel $\theta'(k)$ is some differentiable function,  Eq.~(\ref{eq:kernel}) being an example. We can eliminate $\sigma_{1}(k)$ from these equations to obtain 
\begin{equation}
	\label{eqn:FormBF1}
	\rho_{1}(k) - \int_{-Q}^{Q}dq\;\!\theta_{2}(k,q)\rho_{1}(q)= g_{1}(k),
\end{equation}
where $\theta_{2}(k,q)$ is given by Eq.~(\ref{eq:kernel2}). Note that $\theta_{2}(k,q) = \theta_{2}(q,k)$. Now, consider the coupled set of equations
\begin{subequations}
	\begin{gather}
		\rho_{2}(k) + \frac{1}{\pi}\int_{-B}^{B}dq\;\!\theta'(2k-2q) \sigma_{2}(q)= g_{2}(k),\\
		\sigma_{2}(k) + \frac{1}{\pi}\int_{-Q}^{Q}dq\;\!\theta'(2k-2q) \rho_{2}(q) = 0.
	\end{gather}    
\end{subequations}
Eliminating $\sigma_{2}(k)$, we obtain the equation
\begin{equation}
	\label{eqn:FormBF2}
	\rho_{2}(k) - \int_{-Q}^{Q}dq\;\!\theta_{2}(k,q)\rho_{2}(q)= g_{2}(k).
\end{equation}
Multiplying the Eq.~(\ref{eqn:FormBF1}) by $\rho_{2}(k)$ and the Eq.~(\ref{eqn:FormBF2}) by $\rho_{1}(k)$ and integrating both equations with respect to $k$ we observe that the equations have equal quantities on the left-hand side. Equating the quantities on the right-hand side, we obtain  
\begin{equation}
	\label{eqn:BoseFermi_Integral_P1}
	\int_{-Q}^{Q}dk\;\!\rho_{2}(k) g_{1}(k)= \int_{-Q}^{Q}dk\;\! \rho_{1}(k) g_{2}(k).
\end{equation}
In a similar way, given the coupled equations 
\begin{subequations}
	\begin{gather}
		\rho_{3}(k) + \frac{1}{\pi}\int_{-B}^{B}dq\;\!\theta'(2k-2q) \sigma_{3}(q)= 0,\\
		\sigma_{3}(k) + \frac{1}{\pi}\int_{-Q}^{Q}dq\;\!\theta'(2k-2q) \rho_{3}(q)= g_{3}(k),
	\end{gather}    
\end{subequations}
we have proven the property
\begin{align}
	\label{eqn:BoseFermi_Integral_P2}
	\int_{-Q}^{Q}dk\;\!\rho_{3}(k)g_{1}(k)=\int_{-B}^{B}dk\;\!\sigma_{1}(k)g_{3}(k).
\end{align}
We finally note that given the coupled integral equations
\begin{subequations}
	\begin{gather}
		\rho_{4}(k) + \frac{1}{\pi}\int_{-B}^{B}dq\;\!\theta'(2k-2q) \sigma_{4}(q)= f_{4}(k),\\
		\sigma_{4}(k) + \frac{1}{\pi}\int_{-Q}^{Q}dq\;\!\theta'(2k-2q) \rho_{4}(q)= g_{4}(k),
	\end{gather}    
\end{subequations}
the property
\begin{align}
	\label{eqn:BoseFermi_Integral_P3}
	\int_{-Q}^{Q}dk\;\!\rho_{4}(k)g_{1}(k)={}&\int_{-Q}^{Q}dk\;\! \rho_{1}(k) f_{4}(k)\notag\\
	&+\int_{-B}^{B}dk\;\!\sigma_{1}(k)g_{4}(k)
\end{align}
holds.

\section{The dressed energy equations}\label{sec:chemical_potential_proof}

In this Appendix we consider the dressed energy equations
\begin{subequations}\label{eqn:BF_dressed_T0_app}
\begin{gather}
\mathcal{E}(k) + \frac{1}{\pi}\int_{-B}^{B}dq\;\!\theta'(2k-2q) \varphi(q) = \frac{\hbar^2k^2}{2m} - \bar\mu_{\F}, \\
\varphi(k) + \frac{1}{\pi}\int_{-Q}^{Q} dq\;\!\theta'(2k-2q) \mathcal{E}(q)  = \bar\mu_{\F} - \bar\mu_{\B},
\end{gather}
\end{subequations}
and show that $\bar\mu_{\F}$ and $\bar\mu_{\B}$ coincide with the chemical potentials $\mu_{\F}$ and $\mu_{\B}$, which are used in Eqs.~(\ref{eqn:BF_dressed_T0}).

We begin by defining the functions $\rho_{2}(k)$ and $\sigma_{2}(k)$ by the integral equations 
\begin{subequations}
\label{eqn:BF_mom_2}
\begin{gather}
\rho_{2}(k) + \frac{1}{\pi}\int_{-B}^{B}dq\;\! \theta'(2k-2q) \sigma_{2}(q) dq = \frac{k^2}{2},\\
\sigma_{2}(k) + \frac{1}{\pi}\int_{-Q}^{Q} dq\;\! \theta'(2k-2q) \rho_{2}(q) dq =0.
\end{gather}
\end{subequations}
This enables us to express the functions of Eqs.~(\ref{eqn:BF_dressed_T0_app}) as
\begin{subequations}
\begin{gather}\label{eq:Ekr2kminus}
		\mathcal{E}(k) = \frac{\hbar^2}{m}\rho_{2}(k) - 2\pi\bar\mu_{\F}\rho(k) + 2\pi(\bar\mu_{\F}-\bar\mu_{\B})\tilde{\sigma}(k), \\
		\varphi(k) = \frac{\hbar^2}{m}\sigma_{2}(k) - 2\pi\bar\mu_{\F}\sigma(k) + 2\pi(\bar\mu_{\F}-\bar\mu_{\B})\tilde{\rho}(k),
\end{gather}
\end{subequations}
where we use the functions $\rho(k)$, $\sigma(k)$, $\tilde\rho(k)$, and $\tilde\sigma(k)$ defined by Eqs.~(\ref{eqn:Ground_state_root}) [or its compact form (\ref{eq:BEcompact})] and (\ref{eqn:BF_dual_root}). Since they are the unique solutions, see Appendix \ref{appendix-functional}, from the constraints $\mathcal{E}(Q)=0$ and $\varphi(B)=0$ we obtain the relations
\begin{subequations}
	\label{eqn:BF_dressed_pot}
	\begin{align}
		\bar\mu_{\F} &= \frac{\hbar^2}{2\pi m}\frac{\rho_{2}(Q)\tilde{\rho}_{0}-\sigma_{2}(B)\tilde{\sigma}_{0}}{\rho_{0}\tilde{\rho}_{0}-\sigma_{0}\tilde{\sigma}_{0}},\\
		\bar\mu_{\B} &= \frac{\hbar^2}{2\pi m}\frac{\rho_{2}(Q)(\tilde{\rho}_{0}-\sigma_{0})-\sigma_{2}(B)(\tilde{\sigma}_{0}-\rho_{0})}{\rho_{0}\tilde{\rho}_{0}-\sigma_{0}\tilde{\sigma}_{0}}.
	\end{align}
\end{subequations}

Let us now consider the thermodynamic definition of the chemical potential. It leads to 
\begin{subequations}\label{eqn:chem_pot_def}
	\begin{align}
		\mu_{\F} = \frac{\partial (E_{0}/L)}{\partial n_{\F}} = \left(\frac{\partial Q}{\partial n_{\F}}\frac{\partial (E_{0}/L)}{\partial Q} + \frac{\partial B}{\partial n_{\F}}\frac{\partial (E_{0}/L)}{\partial B}\right),\\
		\mu_{\B} = \frac{\partial (E_{0}/L)}{\partial n_{\B}} = \left(\frac{\partial Q}{\partial n_{\B}}\frac{\partial (E_{0}/L)}{\partial Q} + \frac{\partial B}{\partial n_{\B}}\frac{\partial (E_{0}/L)}{\partial B}\right).
	\end{align}
\end{subequations}
Let us study $\frac{\partial (E_{0}/L)}{\partial Q}$. Using Eq.~(\ref{eq:E0}) and the property (\ref{eqn:BoseFermi_Integral_P1}), we can rewrite it as 
\begin{align}
\frac{\partial (E_{0}/L)}{\partial Q} ={}& \frac{\hbar^2}{2m}
\frac{\partial }{\partial Q}\int_{-Q}^{Q}dk\;\!k^2 \rho(k)\notag\\
={}&\frac{\hbar^2}{2\pi m}\frac{\partial }{\partial Q}\int_{-Q}^{Q}dk\;\!\rho_{2}(k).
\end{align}
Now consider the latter integral,
\begin{equation}
\label{eqn:eq_t1}
\frac{\partial}{\partial Q}\int_{-Q}^{Q}dk\;\!\rho_{2}(k) = 2\rho_{2}(Q)+ \int_{-Q}^{Q}dk\;\!\frac{\partial \rho_{2}}{\partial Q}.
\end{equation}
Differentiating Eqs.~(\ref{eqn:BF_mom_2}) with respect to $Q$, we obtain 
\begin{subequations}
\begin{align}
\mathcal{F}\left[\frac{\partial \rho_{2}}{\partial Q}, \frac{\partial \sigma_{2}}{\partial Q},B\right] = {}&0, \\
\mathcal{F}\left[\frac{\partial \sigma_{2}}{\partial Q}, \frac{\partial \rho_{2}}{\partial Q},Q\right]
={}& -\frac{\rho_{2}(Q)}{\pi}\theta_{+}(k,Q).
\end{align}
\end{subequations}
Thus, using the property (\ref{eqn:BoseFermi_Integral_P2}), we obtain
\begin{align}
\int_{-Q}^{Q}dk\;\!\frac{\partial \rho_{2}}{\partial Q} ={}& -2\rho_{2}(Q)\int_{-B}^{B}dk\;\!\sigma(k)\theta_{+}(k,Q)\notag \\ 
={}& 2\rho_{2}(Q)(2\pi\rho_{0}-1).
\end{align}
In a similar way, we can obtain
\begin{equation}
\int_{-Q}^{Q}dk\;\! \frac{\partial \rho_{2}}{\partial B} = {4\pi\sigma_{2}(B)\sigma_{0}},
\end{equation}
which is needed to calculate $\frac{\partial (E_{0}/L)}{\partial B}$. Collecting all and using the result of Eq.~(\ref{eqn:Matrix_split_RHS2}), we obtain the expression for the chemical potentials,
\begin{subequations}\label{eq:muFmuB}
	\begin{align}
		\mu_{\F} ={}& \frac{\hbar^2}{2\pi m}\frac{\rho_{2}(Q)\tilde{\rho}_{0}-\sigma_{2}(B)\tilde{\sigma}_{0}}{\rho_{0}\tilde{\rho}_{0}-\sigma_{0}\tilde{\sigma}_{0}},\\
		\mu_{\B} ={}& \frac{\hbar^2}{2\pi m}\frac{\rho_{2}(Q)(\tilde{\rho}_{0}-\sigma_{0})-\sigma_{2}(B)(\tilde{\sigma}_{0}-\rho_{0})}{\rho_{0}\tilde{\rho}_{0}-\sigma_{0}\tilde{\sigma}_{0}}. 
	\end{align}
\end{subequations}
A comparison with Eqs.~(\ref{eqn:BF_dressed_pot}) shows that $\mu_{\F}=\bar\mu_{\F}$ and $\mu_{\B} = \bar\mu_{\B}$, which justifies the use of $\mu_{\F}$ and $\mu_{\B}$ in Eqs.~(\ref{eqn:BF_dressed_T0}).

\section{Derivation of the Drude weight matrix}\label{sec:drude_weight_proof2}

In this Appendix we derive the Drude weight matrix. Let us express Eqs.~(\ref{eq:rtst}) in the form
\begin{subequations}\label{eq:rtst1}
\begin{align}
\rho_{t}(k+\delta_2) + \frac{1}{\pi}\int_{-B}^{B}dq\;\! \theta'(2k-2q) \sigma_{t}(q+\delta_2) ={}& \frac{1}{2\pi},\\
\sigma_{t}(k+\delta_1) + \frac{1}{\pi}\int_{-Q}^{Q}dq\;\!\theta'(2k-2q) \rho_{t}(q+\delta_1) ={}& 0.
\end{align}
\end{subequations}
The corresponding ground-state energy is
\begin{align}\label{eq:Et1}
E_t= \frac{\hbar^2 L}{2m}\int_{-Q}^{Q} dk \;\! (k+\delta_1)^2 \rho_{t}(k+\delta_1).
\end{align}
We note that the functions $\rho_t$ and $\sigma_t$ in fact depend on five variables, $(k,\delta_1,\delta_2,Q,B)$. For convenience, we keep only the first one, as in Eqs.~(\ref{eq:rtst1}) and (\ref{eq:Et1}). 

In order to evaluate the Drude weight matrix, it is sufficient to perform a second-order Taylor expansion of the energy (\ref{eq:Et1}) at small $\delta_1$ and $\delta_2$. It leads to
\begin{align}
	\label{eq:Et2}
	E_t={}&E_0+ \frac{\hbar^2 L}{2m}\int_{-Q}^{Q} dk \;\! \biggl[\delta_1 s_1(k) + \delta_2 s_2(k) \notag\\
	&+\frac{1}{2}\delta_1^2 s_{11}(k)+\delta_1\delta_2 s_{12}(k)+\frac{1}{2}\delta_2^2 s_{22}(k)\biggr],
\end{align}
where
\begin{subequations}
\begin{align}
s_1(k)={}&2k \rho(k)+k^2 \rho_t^{(0,1,0)}(k)+k^2 \rho_t^{(1,0,0)}(k),\\
s_2(k)={}&k^2 \rho_t^{(0,0,1)}(k),\\
s_{11}(k)={}&2\rho(k)+k^2\rho_t^{(2,0,0)}(k) +4k \rho_t^{(1,0,0)}(k)\notag\\ &+2k^2 \rho_t^{(1,1,0)}(k)+4k \rho_t^{(0,1,0)}(k)+k^2 \rho_t^{(0,2,0)}(k) ,\\
s_{12}(k)={}&k^2 \rho_t^{(1,0,1)}(k)+2k \rho_t^{(0,0,1)}(k)+k^2 \rho_t^{(0,1,1)}(k),\\
s_{22}(k)={}&k^2\rho_t^{(0,0,2)}(k).
\end{align}
\end{subequations}
Here and in the following we use the notation
\begin{align}
\rho_t^{(i,j,l)}(k)\equiv \frac{\partial^{i+j+l}}{\partial k^i \partial \delta_1^j \partial \delta_2^l}\rho_t(k,\delta_1,\delta_2,\delta_3,Q,B)\bigg{|}_{\delta_1=\delta_2=0}.
\end{align}
Note that for convenience we simply use $\rho_t^{(i,j,l)}(k)$ instead of a more complete form $\rho_t^{(i,j,l)}(k,0,0)$ [or the precise form $\rho_t^{(i,j,l,0,0)}(k,0,0,Q,B)$]. Note also that $\rho_t^{(j,0,0)}(k,0,0)=\frac{\partial^j}{\partial k^j} \rho(k)$, where the latter is defined by Eqs.~(\ref{eqn:Ground_state_root}). For $\sigma_t^{(i,j,l)}(k)$ we use the analogous notation.

The leading-order term $E_0$ of Eq.~(\ref{eq:Et2}) is given by Eq.~(\ref{eq:E0}), which is the energy of the system without the twist, i.e., with periodic boundary conditions. The remaining terms in $E_t$ account for the twisted boundary conditions. They are given by the integrals in Eq.~(\ref{eq:Et2}) and depend on the functions $\rho_t^{(i,j,l)}(k)$ with small $i$, $j$, and $k$. The latter can be found by studying the Taylor expansion of Eqs.~(\ref{eq:rtst1}). Studying the linear order in $\delta_1$, we obtain
\begin{subequations}\label{eq:Frsd1}
\begin{align}
\label{eq:Frsd11}
&\mathcal{F}\left[\rho_t^{(0,1,0)},\sigma_t^{(0,1,0)},B\right]=0,\\
\label{eq:Frsd12}
&\mathcal{F}\left[\sigma_t^{(0,1,0)},\rho_t^{(0,1,0)},Q\right] =-\mathcal{F}\left[\sigma'_k,\rho'_q,Q\right].
\end{align}
\end{subequations}
Here we have used the linearity of the functional and $\rho_t^{(1,0,0)}(k)=\rho'_k(k)$ and $\sigma_t^{(1,0,0)}(k)=\sigma'_k(k)$. Using Eq.~(\ref{eq:rksk2}) and then Eqs.~(\ref{eq:xQoQ}) we obtain
\begin{align}\label{eq:rsd1}
\rho_t^{(0,1,0)}(k)=\frac{\rho_0}{\xi_0}\xi'_Q(k),\quad 
\sigma_t^{(0,1,0)}(k)=\frac{\rho_0}{\xi_0}\omega'_Q(k).
\end{align}
In a similar way, the linear order in $\delta_2$ yields
\begin{subequations}\label{eq:Frsd2}
\begin{align}
&\mathcal{F}\left[\rho_t^{(0,0,1)},\sigma_t^{(0,0,1)},B\right]= -\mathcal{F}\left[\rho'_k,\sigma'_q,B\right],\\
\label{eq:Frsd22}
&\mathcal{F}\left[\sigma_t^{(0,0,1)},\rho_t^{(0,0,1)},Q\right] =0,
\end{align}
\end{subequations}
leading to 
\begin{align}\label{eq:rsd2}
\rho_t^{(0,0,1)}(k)=\frac{\sigma_0}{\omega_0}\xi'_B(k),\quad 
\sigma_t^{(0,0,1)}(k)=\frac{\sigma_0}{\omega_0}\omega'_B(k).
\end{align}
The functions in Eqs.~(\ref{eq:rsd1}) and (\ref{eq:rsd2}) are odd with respect to $k$, which follows from the same property of the right-hand sides of Eqs.~(\ref{eq:Frsd1})  and (\ref{eq:Frsd2}). Since $\rho(k)$ is even and $\rho_t^{(1,0,0)}(k)=\rho'_k(k)$ odd, the terms linear in $\delta_1$ and $\delta_2$ in $E_t$ of Eq.~(\ref{eq:Et2}) vanish.

The Taylor expansion of Eqs.~(\ref{eq:rtst1}) at the order $\delta_1\delta_2$ gives
\begin{subequations}\label{eq:Frsd1d2}
\begin{align}
&\mathcal{F}\left[\rho_t^{(0,1,1)},\sigma_t^{(0,1,1)},B\right]= -\mathcal{F}\left[\rho_t^{(1,1,0)},\sigma_t^{(1,1,0)},B\right],\\
&\mathcal{F}\left[\sigma_t^{(0,1,1)},\rho_t^{(0,1,1)},Q\right]= -\mathcal{F}\left[\sigma_t^{(1,0,1)},\rho_t^{(1,0,1)},Q\right].
\end{align}
\end{subequations}
Here on the right-hand sides we need the expressions obtained by differentiating Eqs.~(\ref{eq:Frsd11}) and (\ref{eq:Frsd22}) with respect to $k$. They are
\begin{subequations}
\begin{align}
&\mathcal{F}\left[\rho_t^{(1,1,0)},\sigma_t^{(1,1,0)},B\right]= \frac{\sigma_t^{(0,1,0)}(B)}{\pi}\theta_+(k,B),\\
&\mathcal{F}\left[\sigma_t^{(1,0,1)},\rho_t^{(1,0,1)},Q\right]= \frac{\rho_t^{(0,0,1)}(Q)}{\pi}\theta_+(k,Q).
\end{align}
\end{subequations}
Using Eqs.~(\ref{eqn:Comb1}) we eventually obtain
\begin{subequations}
\begin{align}
\rho_t^{(0,1,1)}(k)=\frac{\sigma_t^{(0,1,0)}(B)}{\sigma_0} \rho'_B(k)+ \frac{\rho_t^{(0,0,1)}(Q)}{\rho_0} \rho'_Q(k),\\
\sigma_t^{(0,1,1)}(k)=\frac{\sigma_t^{(0,1,0)}(B)}{\sigma_0} \sigma'_B(k)+ \frac{\rho_t^{(0,0,1)}(Q)}{\rho_0} \sigma'_Q(k).
\end{align}
\end{subequations}
The functions $\rho_t^{(0,1,1)}(k)$ and $\sigma_t^{(0,1,1)}(k)$ are even with respect to $k$. The term proportional to $\delta_1 \delta_2$ in $E_t$ of Eq.~(\ref{eq:Et2}) is the integral of $s_{12}(k)$. It is given by
\begin{align}
S_{12}=\int_{-Q}^{Q} dk \;\! \frac{\partial}{\partial k}\left[ k^2 \rho_t^{(0,0,1)}(k)\right]+\int_{-Q}^{Q} dk \;\!k^2 \rho_t^{(0,1,1)}(k).
\end{align}
The first summand of $S$ evaluates to $2Q^2 \sigma_0 \xi'_B(Q)/\omega_0$. The second summand consists of two integrals,
\begin{align}
\int_{-Q}^{Q} dk\;\!k^2\rho_B'(k)={}&4\sigma_0\sigma_2(B),\\
\int_{-Q}^{Q} dk\;\!k^2\rho_Q'(k)={}&2\rho_0[2\rho_2(Q)-Q^2],
\end{align}
which are calculated using Eqs.~(\ref{eqn:BoseFermi_Integral_P1}) and (\ref{eqn:BoseFermi_Integral_P2}). This leads to
\begin{align}\label{eq:S12}
S_{12}=4\rho_0\sigma_{2}(B)\frac{\omega'_Q(B)}{\xi_0}+ 4\sigma_0\rho_2(Q)\frac{\xi'_B(Q)}{\omega_0}.
\end{align}

Let us consider the order $\delta_1^2$ in the Taylor expansion of Eqs.~(\ref{eq:rtst1}). We find
\begin{subequations}\label{eq:rd2sd2}
\begin{align}
\mathcal{F}{}&\left[\rho_t^{(0,2,0)},\sigma_t^{(0,2,0)},B\right]=0,\\ \mathcal{F}{}&\left[\sigma_t^{(0,2,0)},\rho_t^{(0,2,0)},Q\right]= -\mathcal{F}\left[\sigma_t^{(2,0,0)},\rho_t^{(2,0,0)},Q\right]\notag\\ &\phantom{MMMMMMMMiii}-2\mathcal{F}\left[\sigma_t^{(1,1,0)},\rho_t^{(1,1,0)},Q\right] \notag\\
&=-\frac{\rho'_k(Q)+2\rho_t^{(0,1,0)}(Q)}{\pi}\theta_{+}(k,Q)+\frac{\rho_0}{\pi}\frac{d}{dk} \theta_{-}(k,Q).
\end{align}
\end{subequations}
In the last equality we have used Eq.~(\ref{eq:rkkskk2}) and the expression
\begin{align}
\mathcal{F}\left[\sigma_t^{(1,1,0)},\rho_t^{(1,1,0)},Q\right]={}& \frac{\rho_t^{(0,1,0)}(Q)}{\pi} \theta_{+}(k,Q)\notag\\
& - \frac{\rho_0}{\pi}\frac{d}{dk} \theta_{-}(k,Q),
\end{align}
which was obtained  after the differentiation of Eq.~(\ref{eq:Frsd12}) with respect to $k$. A comparison of Eqs.~(\ref{eq:rd2sd2}) with Eqs.~(\ref{eqn:BF_GS_qder}), (\ref{eqn:BF_GS_bder}), and (\ref{eq:xQkoQk}) leads to
\begin{subequations}
\begin{align}
\rho_t^{(0,2,0)}(k)={}&-\frac{\xi'_B(Q)\sigma_0}{\rho_0\omega_0} \rho'_Q(k) - \frac{\omega'_Q(B)\rho_0}{\sigma_0\xi_0}\rho'_B(k)\notag\\
&-\frac{\rho_0}{\xi_0}\xi''_{Qk}(k),\\
\sigma_t^{(0,2,0)}(k)={}&-\frac{\xi'_B(Q)\sigma_0}{\rho_0\omega_0} \sigma'_Q(k) - \frac{\omega'_Q(B)\rho_0}{\sigma_0\xi_0}\sigma'_B(k)\notag\\
&-\frac{\rho_0}{\xi_0}\omega''_{Qk}(k).
\end{align}
\end{subequations}
Here we have used Eq.~(\ref{eqn:BF_Der_rel2_1}).

Applying the relation (\ref{eqn:BoseFermi_Integral_P2}) to Eqs.~(\ref{eqn:BF_exc_spec}) and (\ref{eq:xQoQ}), we find
\begin{align}
\int_{-Q}^{Q}dk\;\!k \xi'_Q(k)={}&-\frac{m\xi_0}{\pi\hbar^2}\int_{-B}^{B}dk\;\!\theta_-(k,Q)\omega(k) \notag\\
={}&\frac{2m}{\hbar^2}\xi_0^2-2Q\xi_0.
\end{align}
This enables us to obtain
\begin{align}
S_{11}={}&\int_{-Q}^{Q}dk s_{11}(k)=\frac{4m}{\hbar^2}\rho_0\xi_0 -S_{12},
\end{align}
where $S_{12}$ is given by Eq.~(\ref{eq:S12}).

Let us consider the order $\delta_2^2$ in the Taylor expansion of Eqs.~(\ref{eq:rtst1}). We obtain
\begin{subequations}\label{eq:rd2sd2qqq}
\begin{align}
\mathcal{F}{}&\left[\rho_t^{(0,0,2)},\sigma_t^{(0,0,2)},B\right]= -\mathcal{F}\left[\rho_t^{(2,0,0)},\sigma_t^{(2,0,0)},B\right]\notag\\ &\phantom{MMMMMMMMiii}-2\mathcal{F}\left[\rho_t^{(1,0,1)},\sigma_t^{(1,0,1)},B\right] \notag\\
&=-\frac{\sigma'_k(B)+2\sigma_t^{(0,0,1)}(B)}{\pi}\theta_{+}(k,B) +\frac{\sigma_0}{\pi}\frac{d}{dk} \theta_{-}(k,B),\\
\mathcal{F}{}&\left[\sigma_t^{(0,0,2)},\rho_t^{(0,0,2)},Q\right]=0.
\end{align}
\end{subequations}
Here we have used Eq.~(\ref{eq:rkkskk1}) and the expression
\begin{align}
	\mathcal{F}\left[\rho_t^{(1,0,1)},\sigma_t^{(1,0,1)},B\right]={}& \frac{\sigma_t^{(0,0,1)}(B)}{\pi} \theta_{+}(k,B)\notag\\
	& - \frac{\sigma_0}{\pi}\frac{d}{dk} \theta_{-}(k,B),
\end{align}
which was obtained  after the differentiation of Eq.~(\ref{eq:Frsd12}) with respect to $k$. A comparison of Eqs.~(\ref{eq:rd2sd2qqq}) with Eqs.~(\ref{eqn:BF_GS_qder}), (\ref{eqn:BF_GS_bder}), and (\ref{eq:xBkoBk}) leads to
\begin{subequations}
\begin{align}
\rho_t^{(0,0,2)}(k)={}&-\frac{\omega'_Q(B)\rho_0}{\sigma_0\xi_0} \rho'_B(k) - \frac{\xi'_B(Q)\sigma_0}{\rho_0\omega_0}\rho'_Q(k)\notag\\
&-\frac{\sigma_0}{\omega_0}\xi''_{Bk}(k),\\
\sigma_t^{(0,0,2)}(k)={}&-\frac{\omega'_Q(B)\rho_0}{\sigma_0\xi_0} \sigma'_B(k) - \frac{\xi'_B(Q)\sigma_0}{\rho_0\omega_0}\sigma'_Q(k)\notag\\
&-\frac{\sigma_0}{\omega_0}\omega''_{Bk}(k).
\end{align}
\end{subequations}
Here we have used Eq.~(\ref{eqn:BF_Der_rel2_2}).

Applying the relation (\ref{eqn:BoseFermi_Integral_P1}) to Eqs.~(\ref{eqn:BF_exc_spec}) and (\ref{eq:xBoB}), we obtain
\begin{align}
	\int_{-Q}^{Q}dk\;\!k \xi'_B(k)=\frac{2m}{\hbar^2}\omega_0^2.
\end{align}
This enables us to find
\begin{align}
	S_{22}={}&\int_{-Q}^{Q}dk s_{22}(k)=\frac{4m}{\hbar^2}\sigma_0\omega_0 -S_{12}.
\end{align}
We have therefore obtained
\begin{align}\label{eq:Etfinal}
E_t={}&E_0+\frac{\hbar^2L}{4m}\left(\delta_1^2 S_{11}+2\delta_1\delta_2 S_{12}+\delta_2^2S_{22}\right)\notag\\
={}&E_0+L\left[\rho_0\xi_0 \delta_1^2+ \sigma_0\omega_0 \delta_2^2\right]-\frac{\hbar^2L}{4m}S_{12}(\delta_1-\delta_2)^2.
\end{align}
At this point we must note that the derivation of the energy difference between $E_t$ and $E_0$ that trivially follows from Eq.~(\ref{eq:Etfinal}) is still incomplete. Namely, it would make physical sense to subtract the two energies for the two systems with two different sets of boundary conditions only if they are calculated for the same number of particles. Since $\delta_1,\delta_2\propto 1/L$, see Eqs.~(\ref{eqn:delta_value}), the energy difference scales as $1/L$ and represents a subleading term in the energy. We should thus calculate the densities of particles in the two systems within the same accuracy.

The densities of all particles $n_t$ and of bosons $n_{\B t}$ in the case of twisted boundary conditions follow from
\begin{subequations}
\begin{gather}
n_t=\int_{-Q}^{Q}dk\;\!\rho_t(k+\delta_1),\\
n_{\B t}=\int_{-B}^{B}dk\;\!\sigma_t(k+\delta_2).
\end{gather}
\end{subequations}
Its evaluation simple once we have derived all the steps to calculate $E_t$. Expanding in the Taylor series at small $\delta_1$ and $\delta_2$ to the second (i.e., leading) order, we obtain the general forms
\begin{subequations}
\begin{gather}
n_t=n+\frac{1}{2}n_{11}\delta_1^2+ n_{12}\delta_1\delta_2 +\frac{1}{2}n_{22}\delta_2^2,\\
n_{\B t}=n_{\B}+\frac{1}{2}\tilde n_{11}\delta_1^2+ \tilde n_{12}\delta_1\delta_2 +\frac{1}{2}\tilde n_{22}\delta_2^2.
\end{gather}
\end{subequations}
As expected, the leading-order terms are $n$ and $n_{\B}$. The terms linear in $\delta_1$ and $\delta_2$ do not exist as the corresponding integrands are odd and give zero after the integration. 

The terms proportional to $\delta_1 \delta_2$ are
\begin{subequations}
\begin{align}
n_{12}=\int_{-Q}^{Q}dk\;\!\left[\rho_t^{(0,1,1)}(k)+\rho_t^{(1,0,1)}(k) \right],\\
\tilde n_{12}=\int_{-B}^{B}dk\;\!\left[\sigma_t^{(0,1,1)}(k)+\sigma_t^{(1,1,0)}(k) \right].
\end{align}
\end{subequations}
After evaluation we obtain 
\begin{subequations}
\begin{align}
n_{12}={}&4\pi\rho_0\sigma_0\left[\frac{\omega'_Q(B)}{\xi_0}+ \frac{\xi'_B(Q)}{\omega_0}\right],\\
\tilde n_{12}={}&4\pi\rho_0\tilde\rho_0\frac{\omega'_Q(B)}{\xi_0}+ 4\pi\sigma_0\tilde\sigma_0 \frac{\xi'_B(Q)}{\omega_0}.
\end{align}
\end{subequations}
Here we have used
\begin{subequations}
\begin{align}
\int_{-Q}^{Q} dk\;\!\rho_B'(k)={}&\frac{\partial n}{\partial B}=4\pi\sigma_0^2,\\
\int_{-Q}^{Q} dk\;\!\rho_Q'(k)={}&\frac{\partial n}{\partial Q} -2\rho_0= 4\pi\rho_0^2-2\rho_0,\\
\int_{-B}^{B} dk\;\!\sigma_B'(k)={}&\frac{\partial n_{\B}}{\partial B}-2\sigma_0=4\pi\sigma_0\tilde \rho_0-2\sigma_0,\\
\int_{-B}^{B} dk\;\!\sigma_Q'(k)={}&\frac{\partial n_{\B}}{\partial Q}= 4\pi\rho_0\tilde\sigma_0.
\end{align}
\end{subequations}
The terms proportional to $\delta_2^2$ are
\begin{align}
n_{22}={}&\int_{-Q}^{Q}dk\;\!\rho_t^{(0,0,2)}(k)=-n_{12},\\
\tilde n_{22}={}&\int_{-B}^{B}dk\;\! \left[\sigma_t^{(0,0,2)}(k)+\sigma_t^{(2,0,0)}(k)+ 2\sigma_t^{(1,0,1)}(k) \right]\notag\\
={}&-\tilde n_{12}.
\end{align}
The terms proportional to $\delta_1^2$ are
\begin{subequations}
\begin{align}
n_{11}={}&\int_{-Q}^{Q}dk\;\!\left[\rho_t^{(0,2,0)}(k)+\rho_t^{(2,0,0)}(k)+ 2\rho_t^{(1,1,0)}(k) \right]\notag\\
={}&-n_{12},\\
\tilde n_{11}={}&\int_{-B}^{B}dk\;\! \sigma_t^{(0,2,0)}(k)=-\tilde n_{12}.
\end{align}
\end{subequations}
We thus eventually obtain
\begin{subequations}
\begin{gather}\label{eq:nt}
n_t=n-\frac{1}{2} n_{12}(\delta_1-\delta_2)^2,\\
n_{\B t}=n_{\B}-\frac{1}{2} \tilde n_{12}(\delta_1-\delta_2)^2.
\end{gather}
We can therefore observe that $n_t$ and $n$ as well as $n_{\B t}$ and $n_{\B}$ differ once we account for the subleading terms.

\end{subequations}
Using Eqs.~(\ref{eq:muFmuB}), we find that the energy (\ref{eq:Etfinal}) can be equivalently expressed as
\begin{align}\label{eq:end}
E_t={}&E_0+L\mu_{\B}(n_{\B t}-n_{\B})+L \mu_{\F}(n_{\F t}-n_{\F})\notag\\
&+L\left[\rho_0\xi_0 \delta_1^2+ \sigma_0\omega_0 \delta_2^2\right].
\end{align}
Here $n_{\F t}=n_t-n_{\B t}$ and $n_{\F}=n-n_{\B}$ are the densities of fermions for the two sets of boundary conditions. Equation (\ref{eq:end}) is our final expression with a clear interpretation. It shows that the energies of the two systems with different boundary conditions differ by the trivial terms that account for the particle number difference and the nontrivial term given by the second line of Eq.~(\ref{eq:end}). The nontrivial term is the one that determines the Drude weight matrix, which thus follows from the energy difference calculated for the same densities,
\begin{align}\label{eq:E38}
E_{t}-E_{0} ={}& \frac{2\pi\hbar}{L}\Bigl\{\bigl[v_{1}(\rho_{0}-\tilde{\sigma}_{0})^2 + v_{2}(\sigma_{0}-\tilde{\rho}_{0})^2\bigr]\varphi_{\F}^2\notag\\ &+2\bigl[v_{1}\tilde{\sigma}_{0}(\rho_{0}-\tilde{\sigma}_{0}) + v_{2}\tilde{\rho}_{0}(\sigma_{0}-\tilde{\rho}_{0})\bigr]\varphi_{\F}\varphi_{\B} \notag\\
&+\bigl[v_{1}\tilde{\sigma}_{0}^2 + v_{2}\tilde{\rho}_{0}^2\bigr]\varphi_{\B}^2\Bigr\}.
\end{align}
Here we have used Eqs.~(\ref{eqn:delta_value}) to express  $\delta_{1}$ and $\delta_{2}$ in terms of the twisting angles. 

We note that if we used the twisted boundary conditions with one phase in Eq.~(\ref{eq:TWBC}), $\varphi=\varphi_{\F}=\varphi_{\B}$, the resulting energy difference (\ref{eq:E38}) would  significantly simplify to
\begin{align}
E_{t}-E_{0} ={}& \frac{2\pi\hbar}{L}\left(v_1 \rho_0^2+v_2\sigma_0^2\right)\varphi^2\notag\\
={}&\frac{\hbar^2 n}{2mL}\varphi^2.
\end{align}
Here we have used the relation (\ref{eqn:Gal_inv_rel_1}) that arises due to Galilean invariance. The resulting Drude weight is just one matrix element given by Eq.~(\ref{eq:Drudematrixelements}), which is 
\begin{align}\label{eq:E40}
\mathcal{D}=\frac{\pi \hbar n}{m}.
\end{align}
Alternatively, for $\varphi=\varphi_{\F}=\varphi_{\B}$, we have $\mathcal{D}=\mathcal{D}_{\FF}+2\mathcal{D}_{\FB}+\mathcal{D}_{\BB}$ that again leads to the result (\ref{eq:E40}) upon using the sum rules (\ref{eq:Dsumrules}). The result (\ref{eq:E40}) is consistent with the recent observation made in Ref.~\cite{gohmann_ballistic_2025}.

Let us also evaluate the momentum in the system with twisted boundary conditions using
\begin{align}\label{eq:Pt1}
P_t=\hbar L\int_{-Q+\delta_1}^{Q+\delta_1} dk\;\! k \rho_t(k).
\end{align}
Performing the expansion to the linear order in $\delta_1$ and $\delta_2$ we obtain
\begin{align}
P_t=\hbar L\int_{-Q}^{Q}dk\;\! (k+\delta_1)\bigl\{\rho(k)+\delta_2\rho_t^{(0,0,1)}(k)\notag\\
+\delta_1[\rho_t^{(1,0,0)}(k)+ \rho^{(0,1,0)}(k)]+\ldots\bigr\}.
\end{align}
Previously derived expressions in this Appendix enable us to eventually find
\begin{align}
P_t=\frac{2mL}{\hbar}(\rho_0\xi_0\delta_1+\sigma_0\omega_0\delta_2).
\end{align}
Using Eqs.~(\ref{eqn:delta_value}) to express  $\delta_{1}$ and $\delta_{2}$ in terms of the twisting angles, we recover the result (\ref{eq:Pt}). Therefore Eq.~(\ref{eq:Pt1}) leads to the same result as the discrete expression of Eq.~(\ref{eq:Pt}) in the leading order for small angles, which should be the case.

\section{Alternative derivation of the Drude weight matrix}\label{sec:drude_weight_proof}

Here we provide an alternative derivation to the one of Appendix~\ref{sec:drude_weight_proof2} of the Drude weight matrix. Let us consider the dressed energy equations for the system with twisted boundary conditions that are given by 
\begin{subequations}\label{eq:Etphit}
\begin{align}
\mathcal{E}_{t}(k) + \frac{1}{\pi}\int_{-B+\delta_{2}}^{B+\delta_{2}}dq\;\!\theta'(2k-2q) \varphi_{t}(q) &= \frac{\hbar^2 k^2}{2m}-\mu_{\F},\\
\varphi_{t}(k) + \frac{1}{\pi}\int_{-Q+\delta_{1}}^{Q+\delta_{1}}dq\;\!\theta'(2k-2q) \mathcal{E}_{t}(q) &= \mu_{\F}-\mu_{\B}.
\end{align}
\end{subequations}
Here $\mu_{\F}$ and $\mu_{\B}$ are the chemical potentials defined by Eqs.~(\ref{eqn:chem_pot_def}). Note that the functions $\mathcal{E}_{t}$ and $\varphi_{t}$ actually depend on five variables, $(k,\delta_1,\delta_2,Q,B)$, but for our present derivation we only need $(k,\delta_1,\delta_2)$. For convenience we keep only the first one in Eqs.~(\ref{eq:Etphit}). In the case of zero twisting angles, $\delta_1=\delta_2=0$, Eqs.~(\ref{eq:Etphit}) are identical to Eqs.~(\ref{eqn:BF_dressed_T0}). Thus 
\begin{align}\label{eq:equality}
\mathcal{E}_t(k)=\mathcal{E}(k),\quad \varphi_t(k)=\varphi(k),
\end{align}
where $\delta_1=\delta_2=0$ should be used in the left-hand sides. The functions entering Eq.~(\ref{eq:equality}) are even with respect to $k$. Equation (\ref{eq:equality}) implies 
\begin{align}\label{eq:edgeproperty}
\mathcal{E}_{t}(\pm Q)=\varphi_{t}(\pm B)=0\quad\textrm{for}\quad  \delta_{1}=\delta_{2}=0.
\end{align}

Consider the generalized energy defined via
\begin{equation}
\label{eqn:Gibbs_twisted}
    G_{t}(\delta_1,\delta_2) = \frac{L}{2\pi}\int_{-Q+\delta_{1}}^{Q+\delta_{1}}dk\;\!\mathcal{E}_{t}(k).
\end{equation}
Here we omitted explicit dependence on $Q$ and $B$ in the arguments of $G_t$. Let us analogously define
\begin{equation}
\label{eqn:Gibbs_normal}
G = \frac{L}{2\pi}\int_{-Q}^{Q}dk\;\!\mathcal{E}(k).
\end{equation}
We want to calculate the difference $G_t(\delta_1,\delta_2)-G$ at small $\delta_1$ and $\delta_2$. The corresponding Taylor series of the second order is given by 
\begin{align}
G_{t}-G = {}& \delta_{1}G_t^{(1,0)} + \delta_{2}G_t^{(0,1)}\notag\\ 
&+ \frac{1}{2}\begin{pmatrix} \delta_{1} & \delta_{2}
\end{pmatrix}\begin{pmatrix}
G_t^{(2,0)} & G_t^{(1,1)}\\
G_t^{(1,1)} & G_t^{(0,2)}
\end{pmatrix}\begin{pmatrix}
\delta_{1} \\
\delta_{2}
\end{pmatrix}+\ldots.
\end{align}
All partial derivatives with respect to $\delta_1$ and $\delta_2$ are evaluated at $\delta_1=\delta_2=0$. Consider the first partial derivatives of $G_t$. They are
\begin{subequations}
\begin{align}
G_{t}^{(1,0)} &= \frac{L}{2\pi}\int_{-Q}^{Q}dk\;\!\mathcal{E}_{t}^{(0,1,0)}(k),\\
G_{t}^{(0,1)} &= \frac{L}{2\pi}\int_{-Q}^{Q}dk\;\! \mathcal{E}_{t}^{(0,0,1)}(k).
\end{align}
\end{subequations}
where we used the boundary conditions (\ref{eq:edgeproperty}). The integral equations for the first derivatives $\mathcal{E}_t^{(0,1,0)}(k)$ and $\mathcal{E}_t^{(0,0,1)}(k)$ (at $\delta_{1}=\delta_{2}=0$) follow from Eqs.~(\ref{eq:Etphit}). They are
\begin{subequations}
\begin{align}
\mathcal{F}\left[\mathcal{E}_t^{(0,1,0)},\varphi_{t}^{(0,1,0)},B\right]&=0,\\
\mathcal{F}\left[\varphi_{t}^{(0,1,0)},\mathcal{E}_t^{(0,1,0)},Q\right]&=-\frac{\mathcal{E}_{t}(Q)}{\pi}\theta_{-}(k,Q)=0,
\end{align}
\end{subequations}
and 
\begin{subequations}
\begin{align}
\mathcal{F}\left[\mathcal{E}_t^{(0,0,1)},\varphi_{t}^{(0,0,1)},B\right]&=-\frac{\varphi_{t}(B)}{\pi}\theta_{-}(k,B)=0,\\
\mathcal{F}\left[\varphi_{t}^{(0,0,1)},\mathcal{E}_t^{(0,0,1)},Q\right]&=0.
\end{align}
\end{subequations}
Here we have used Eq.~(\ref{eq:edgeproperty}). We therefore find
\begin{subequations}\label{eq:Et010}
\begin{gather}
\mathcal{E}_{t}^{(0,1,0)}(k)=0,\quad\mathcal{E}_{t}^{(0,0,1)}(k)=0, \\
\varphi_{t}^{(0,1,0)}(k)=0,\quad\varphi_{t}^{(0,0,1)}(k)=0,
\end{gather}
\end{subequations}
and thus $G_t^{(1,0)}=G_t^{(0,1)}=0$.

The second-order correction to $G_t$ is determined by the second derivatives
\begin{subequations}\label{eq:secGt}
\begin{align}
G_t^{(2,0)}={}& \frac{L}{2\pi}\left[2\mathcal{E}'_{k}(Q)+\int_{-Q}^{Q}dk\;\! \mathcal{E}_{t}^{(0,2,0)}(k)\right], \\
G_t^{(0,2)} ={}& \frac{L}{2\pi}\int_{-Q}^{Q}dk\;\! \mathcal{E}_{t}^{(0,0,2)}(k),\\
G_t^{(1,1)}={}& \frac{L}{2\pi}\int_{-Q}^{Q}dk\;\! \mathcal{E}_{t}^{(0,1,1)}(k).
\end{align}
\end{subequations}
The functions that enter Eqs.~(\ref{eq:secGt}) are obtained by differentiating Eqs.~(\ref{eq:Etphit}) with respect to $\delta_1$ and $\delta_2$ followed by setting $\delta_{1}=\delta_{2}=0$. The resulting integral equations are  
\begin{subequations}
\begin{align}
\mathcal{F}\left[\mathcal{E}_{t}^{(0,2,0)},\varphi_{t}^{(0,2,0)},B\right] = {}& 0,\\
\mathcal{F}\left[\varphi_{t}^{(0,2,0)},\mathcal{E}_{t}^{(0,2,0)},Q\right] = {}& -\frac{\mathcal{E}'_{k}(Q)}{\pi}\theta_{+}(k,Q),
\end{align}
\end{subequations}
for the second derivative with respect to $\delta_1$,
\begin{subequations}
\begin{align}
\mathcal{F}\left[\mathcal{E}_{t}^{(0,0,2)},\varphi_{t}^{(0,0,2)},B\right] = {}& - \frac{\varphi'_{k}(B)}{\pi}\theta_{+}(k,B), \\
\mathcal{F}\left[\varphi_{t}^{(0,0,2)},\mathcal{E}_{t}^{(0,0,2)},Q\right] = {}& 0,
\end{align}
\end{subequations}
for the second derivative with respect to $\delta_2$, and
\begin{subequations}
\label{eqn:Twisted_mixed}
\begin{align}
\mathcal{F}\left[\mathcal{E}_{t}^{(0,1,1)},\varphi_{t}^{(0,1,1)},B\right] ={}& -\frac{\varphi_{t}^{(0,1,0)}(B)}{\pi} \theta_{+}(k,B)=0,\\
\mathcal{F}\left[\varphi_{t}^{(0,1,1)},\mathcal{E}_{t}^{(0,1,1)},Q\right] ={}& -\frac{\mathcal{E}_{t}^{(0,0,1)}(Q)}{\pi}\theta_{+}(k,Q)=0, 
\end{align}
\end{subequations}
for the mixed derivative, where we have used Eqs.~(\ref{eq:Et010}). Therefore, 
Eqs.~(\ref{eqn:Twisted_mixed}) imply $\mathcal{E}_{t}^{(0,1,1)}(k)= \varphi_{t}^{(0,1,1)}(k)=0$, which means $G_t^{(1,1)}=0$. The remaining integrals are evaluated using the properties (\ref{eqn:BoseFermi_Integral_P1}) and (\ref{eqn:BoseFermi_Integral_P2}), leading to 
\begin{subequations}
\begin{align}
\int_{-Q}^{Q}dk\;\!\mathcal{E}_{t}^{(0,0,2)} (k) ={}& -2 \mathcal{E}'_{k}(Q)\int_{-B}^{B}dk\;\!\sigma(k)\theta_{+}(k,Q)\notag\\
={}& 4\pi\rho_{0}\xi_{0}-2\xi_0,\\   
\int_{-Q}^{Q}dk\;\!\mathcal{E}_{t}^{(0,2,0)}(k) ={}&  -2\varphi'_{k}(B)\int_{-Q}^{Q}dk\;\!\rho(k)\theta_{+}(k,B)\notag \\
={}& 4\pi \sigma_{0}\omega_{0}.
\end{align}
\end{subequations}
Here we have used Eq.~(\ref{eq:equality}) and $\mathcal{E}'_{k}(k)=\xi(k)$, $\varphi'_{k}(k)=\omega(k)$. Combining the above results, we obtain
\begin{align}
G_{t}-G = L(\xi_{0}\rho_{0}\delta_{1}^2+\omega_{0}\sigma_{0}\delta_{2}^2)+\ldots,
\end{align}
where the ellipsis denotes higher-order corrections in $\delta_1$ and $\delta_2$.

There remains to find the relation between $G_t$ and $E_t$ that are defined by Eqs.~(\ref{eqn:Gibbs_twisted}) and (\ref{eq:Et}), respectively. Using the property (\ref{eqn:BoseFermi_Integral_P3}) for the wanted relation we find
\begin{equation}
G_{t} = E_{t}-L\mu_{\F}n_{\F t}-L\mu_{\B}n_{\B t},
\end{equation}
and analogously
\begin{equation}\label{eq:GE0}
G = E_0-L\mu_{\F}n_{\F}-L\mu_{\B}n_{\B},
\end{equation}
where the latter directly follows from Eq.~(\ref{eq:Ekr2kminus}). Then the relation (\ref{eq:end}) directly follows. We have therefore achieved our goal and derived the energy difference $E_t-E_0$ starting from the dressed energy equations (\ref{eq:Etphit}).

\section{Various derivatives of Eqs.~(\ref{eqn:Ground_state_root}) and (\ref{eqn:BF_exc_spec})}\label{appendix:last}

Here we collect various relations obtained by directly differentiating Eqs.~(\ref{eqn:Ground_state_root}) and (\ref{eqn:BF_exc_spec}) that were not stated elsewhere. They are written in the compact notation. 

The first derivative with respect to $k$ of Eqs.~(\ref{eqn:Ground_state_root}) is
\begin{subequations}
\label{eq:rksk}
\begin{align}
\label{eq:rksk1}
\mathcal{F}\left[\rho'_k,\sigma'_q,B\right]=\frac{\sigma_0}{\pi}\theta_{-}(k,B),\\
\label{eq:rksk2}
\mathcal{F}\left[\sigma'_k,\rho'_q,Q\right]=\frac{\rho_0}{\pi}\theta_{-}(k,Q),
\end{align}
\end{subequations}
and the second derivative with respect to $k$ is 
\begin{subequations}
\label{eq:rkkskk}
\begin{align}
\label{eq:rkkskk1}
\mathcal{F}\left[\rho''_{kk},\sigma''_{qq},B\right]=\frac{\sigma'_k(B)}{\pi}\theta_{+}(k,B) +\frac{\sigma_0}{\pi}\frac{d}{dk} \theta_{-}(k,B),\\
\label{eq:rkkskk2}
\mathcal{F}\left[\sigma''_{kk},\rho''_{qq},Q\right]=\frac{\rho'_k(Q)}{\pi}\theta_{+}(k,Q) +\frac{\rho_0}{\pi}\frac{d}{dk} \theta_{-}(k,Q).
\end{align}
\end{subequations}
The first derivative of Eqs.~(\ref{eqn:BF_exc_spec}) with respect to $Q$ is
\begin{subequations}
\label{eq:xQoQ}
\begin{align}
\label{eq:xQoQ1}
\mathcal{F}[\xi'_{Q},\omega '_{Q},B] &=0, \\
\label{eq:xQoQ2}
\mathcal{F}[\omega '_{Q},\xi'_{Q},Q] &=-\frac{\xi_0}{\pi}\theta_{-}(k,Q),
\end{align}
\end{subequations}
and the first derivative with respect to $B$ is
\begin{subequations}
\label{eq:xBoB}
\begin{align}
\mathcal{F}[\xi'_{B},\omega '_{B},B] &=-\frac{\omega_0}{\pi}\theta_{-}(k,B), \\
\mathcal{F}[\omega '_{B},\xi'_{B},Q] &=0.
\end{align}
\end{subequations}
Finally, the second mixed derivative of Eqs.~(\ref{eqn:BF_exc_spec}) with respect to $Q$ and $k$ is
\begin{subequations}
\label{eq:xQkoQk}
\begin{align}
\label{eq:xQkoQk1}
\mathcal{F}[\xi''_{Qk},\omega''_{Qq},B] &=\frac{\omega'_Q(B)}{\pi}\theta_{+}(k,B), \\
\label{eq:xQkoQk2}
\mathcal{F}[\omega''_{Qk},\xi''_{Qq},Q] &= \frac{\xi'_Q(Q)}{\pi}\theta_{+}(k,Q) -\frac{\xi_0}{\pi}\frac{d}{dk}\theta_{-}(k,Q),
\end{align}
\end{subequations}
while the second mixed derivative with respect to $B$ and $k$ reads
\begin{subequations}
\label{eq:xBkoBk}
\begin{align}
\mathcal{F}[\xi''_{Bk},\omega''_{Bq},B] &=\frac{\omega'_B(B)}{\pi}\theta_{+}(k,B)-\frac{\omega_0}{\pi}\frac{d}{dk}\theta_{-}(k,B), \\
\mathcal{F}[\omega''_{Bk},\xi''_{Bq},Q] &=\frac{\xi'_B(Q)}{\pi}\theta_{+}(k,Q).
\end{align}
\end{subequations}
We note the differentiation rule for the operator,
\begin{align}
&\frac{\partial}{\partial k} \mathcal{F}[f(k),g(q),B]=\mathcal{F}\left[\frac{\partial f}{\partial k},\frac{\partial g}{\partial q},B\right]\notag\\
&-\frac{1}{\pi}\left[g(B)\theta'(2k-2B)-g(-B)\theta'(2k+2B)\right].
\end{align}
Depending on the parity of $g(q)$, the last term can be recast in terms of $\theta_{-}(k,B)$ for even and in terms of $\theta_{+}(k,B)$ for odd $g(q)$.

%\bibliography{bibliog}


\end{document}